\documentclass[showpacs,aps,pra,twocolumn,10pt,superscriptaddress]{revtex4-2}

\usepackage{amsmath,graphicx,color}

\begin{document}

\title{Spin fluctuations in the dissipative phase transitions of the quantum Rabi model}

\author{Jiahui Li}
\affiliation{Beijing Computational Science Research Center, Beijing 100193, China}

\author{Rosario Fazio}
\affiliation{The Abdus Salam International Center for Theoretical Physics (ICTP), Strada Costiera 11, 34151 Trieste, Italy}
\affiliation{Dipartimento di Fisica, Universit\`a di Napoli Federico II, Monte S. Angelo, I-80126 Napoli, Italy}

\author{Yingdan Wang}
\email[]{yingdan.wang@itp.ac.cn}
\affiliation{Institute of Theoretical Physics, Chinese Academy of Sciences, Beijing 100190, China}
\affiliation{School of Physical Sciences, University of Chinese Academy of Sciences, Beijing 100049, China}

\author{Stefano Chesi}
\email[]{stefano.chesi@csrc.ac.cn}
\affiliation{Beijing Computational Science Research Center, Beijing 100193, China}
\affiliation{Department of Physics, Beijing Normal University, Beijing 100875, China}

\begin{abstract}
We investigate the dissipative phase transitions of the anisotropic quantum Rabi model with cavity decay and demonstrate that large spin fluctuations persist in the stationary state, having important consequences on the phase diagram and the critical properties. In the second-order phase transition to the superradiant phase, there is a significant suppression of the order parameter and the appearance of non-universal factors, which directly reflect the spin populations. Furthermore, upon entering a parameter regime where mean-field theory predicts a tricritical phase, we find a first-order phase transition due to the unexpected collapse of superradiance. An accurate and physically transparent description going beyond mean-field theory is established by combining exact numerical simulations, the cumulant expansion, and analytical approximations based on reduced master equations and an effective equilibrium theory. Our findings, compared to the conventional thermodynamic limit of the Dicke model, indicate a general tendency of forming extreme non-equilibrium states in the single-spin system, thus have broad implications for dissipative phase transitions of few-body systems.
\end{abstract}

\maketitle

\section{Introduction}\label{intro}

Coupling qubits to discrete bosonic modes is at the heart of quantum information processing and quantum simulators. It allows to realize tunable long-range interactions, useful to implement quantum gates or desired many-body Hamiltonians. Besides cavity electrodynamics \cite{HarocheRMP2001,EsslingerRMP2001}, remarkable collective phenomena such as dissipative phase transitions \cite{FitzpatrickPRX2017,CollodoPRL2019}, many-body localization \cite{SmithNPhys2016,RubemNPhys2021}, time crystals \cite{MonroeNature2017}, and dynamical phase transitions \cite{Shen2017PRL,MonroeNature2017dpt} were explored in trapped ions \cite{MonroeRMP2021} and circuit QED \cite{BlaisRMP2021} platforms, where qubit interactions are mediated by vibrational or cavity modes. Coupling multiple spin qubits to superconducting resonators has also recently witnessed substantial progress \cite{2020_Nature_Petta,2022_PRX_Vandersypen}. 

Among these spin-boson systems, the Dicke model is one of the most actively investigated paradigms of light-matter interactions~\cite{Dicke1954,kirton2019introduction}. It describes $N$ qubits collectively coupled to a single-mode cavity which, for a sufficiently strong interaction, undergoes a transition to a symmetry-broken superradiant state. This superradiant phase transition can occur both in equilibrium~\cite{hepp1973superradiant,HioePRA1973} and in the driven-dissipative case~\cite{DimerPRA2007,kirton2017suppressing}, and has been observed in various cold atoms and trapped ions experiments~\cite{BaumannNature2010, KlinderPNAS2015, ZhiqiangOptica2017, ParkinsPRA2018, SafaniNainiPRL2018}. In recent years, the strictly related concept of a superradiant phase transition of the Rabi model~\cite{levine2004entanglement,HinesPRA2004, BakemeierPRA2012,ashhab2013superradiance} (with $N=1$) has also attracted considerable attention~\cite{hwang2015quantum, puebla2017probing, liu2017universal, hwang2018dissipative}. In the absence of a conventional thermodynamic limit, such transition takes place for a large value of the ratio between the atomic and optical frequencies. The idea has been generalized to a variety of setups, including models with multiple sites and central-spin systems~\cite{PhysRevLett.117.123602,PhysRevLett.124.040404,PhysRevA.101.063843,PhysRevA.107.013714,PhysRevLett.127.063602,PhysRevLett.129.183602,PhysRevLett.130.043602,PhysRevA.107.033702,PhysRevLett.130.210404}. Remarkably, the single-atom phase transition has been recently implemented with a trapped ion coupled to its vibrational motion \cite{puebla2017probing,KimPRX2018,DuanNatComm2021}, and has been shown to have great value for quantum metrology applications~\cite{2020_PRL_Paris_metrology,2021_PRL_sensing,2022_PRXQuantum_sensing,2023_SChina_sensing}.

In this article, we take a fresh look at the dissipative phase transition of the Rabi model~\cite{hwang2018dissipative} and demonstrate the crucial role played by spin fluctuations in correctly understanding the properties of the system. This is an unexpected finding within this class of models, for which the semiclassical treatment usually becomes exact in the appropriate thermodynamic limit. Here, instead, spin fluctuations survive in the stationary state and significantly suppress the order parameter of the superradiant phase. Their effect is already large in the isotropic case of balanced co- and counter-rotating interactions, when the spin is driven to an infinite-temperature state, but can be further enhanced if counter-rotating terms dominate. Most dramatically, we observe a sudden collapse of superradiance in a tristable region of the phase diagram, where a superradiant state would be allowed within mean-field theory. The disappearance of superradiance is due to an effective incoherent pumping process of the spin, which ultimately leads to population trapping in the normal state and gives rise to a first-order transition between the normal and superradiant phases. Thus, spin fluctuations induce profound modifications in the structure of the phase diagram. 


The need of a careful treatment of quantum fluctuations is already clear from mean-field theory, where multi-stability is found in the whole phase diagram. For the Dicke model, bistability is only allowed in limited regions of parameters~\cite{KeelingPRL2010,Soriente2018,Stitely2020}, in the experimentally relevant~\cite{ZhiqiangOptica2017} case of finite anisotropy (i.e., unbalanced co- and counter-rotating interactions~\cite{DimerPRA2007,baksic2014controlling,xie2014anisotropic,YangPRA2017,shen2017quantum,liu2017universal,Soriente2018,Stitely2020}). Here, instead, the normal state with an excited qubit is always stable in the appropriate `thermodynamic' limit. Bistability persists in the isotropic model and a tristable phase appears at large anisotropy. The fate of multistablity is determined by quantum jumps between mean-field solutions, thus requires an analysis of the stationary state based on the full master equation.

After first establishing the above picture within mean-field theory, we perform a preliminary study of quantum fluctuations based on the cumulant expansion in the normal phase. We then provide a comprehensive and physically transparent treatment of the stationary state and derive the corresponding phase diagram. The stationary state of the normal phase is already non-trivial, as it is determined by a competition between effective spin relaxation and incoherent pumping. As mentioned, these processes bring the qubit to an infinite-temperature state in the balanced model, and lead to population inversion when counter-rotating terms dominate. In the superradiant phase, the relaxation and pumping processes become more involved, but are still amenable of analytical treatment and their effect can be summarized by relatively simple transition diagrams (see Figs.~\ref{schematics_rates_1} and \ref{schematics_rates_2}). The dissipative dynamics leads to a significant suppression of the order parameter from an ideal supperradiant state (see, e.g., Fig.~\ref{fig5-photon}), due to the finite population in the high-energy normal state. Identifying the structure of relevant transitions between semiclassical states not only allows us to describe accurately the superradiant order parameter, but also leads to a clear explanation for the absence of superradiance in the tricritical region (see Fig.~\ref{schematics_rates_2}).

Furthermore, we show that at the second-order phase transition an effective equilibrium theory originally developed for the isotropic Dicke model~\cite{dalla2013keldysh} can be extended to the anisotropic Rabi model. This approach allows us to derive accurate analytic formulas for the scaling functions and the non-universal factors relating the two (Rabi and Dicke) models. In particular, for the isotropic case the exact value of the non-universal factor is 1/2, which agrees with previous numerical findings~\cite{hwang2018dissipative}. While the precise origin of these non-universal factors has not been fully appreciated, our analysis shows that they coincide with the spin populations. This simple interpretation links in a direct manner these prefactors of the scaling behavior to the nontrivial spin dynamics taking place in the Rabi model.

The structure of our paper is as follows: In Sec.~\ref{model} we define the model and in Sec.~\ref{mean_field} we obtain the phase diagram within mean-field theory. In Sec.~\ref{sec:finite_eta} we treat the leading quantum corrections induced by a finite value of $\eta$. In particular, in Sec.~\ref{cumulant} we use the cumulant expansion to capture quantum fluctuations in the normal phase. An alternative approach based on an effective master equation is introduced in Sec.~\ref{effective-me}, which is later generalized in Sec.~\ref{sec:second_order_transition} to the superradiant phase and in Sec.~\ref{sec:Cphase} to the tricritical phase. This allows us to characterize the stationary state in general, and derive the corresponding phase diagram. Finally, in Sec.~\ref{finite frequency scaling} we show that a modified effective equilibrium theory is able to  describe accurately the critical scaling of the second-order phase transition. Our conclusions are presented in Sec.~\ref{concl}.

\section{Model}\label{model}

The Hamiltonian of the anisotropic Rabi model reads:
\begin{align}
\label{eq1}
\hat{H}=&\omega_c\hat{a}^{\dagger}\hat{a}+\frac{\Omega}{2}\hat{\sigma}_z - \lambda_-\left(\hat{a} \hat{\sigma}_++ \hat{a}^\dag \hat{\sigma}_- \right) \nonumber \\
& - \lambda_+ \left(\hat{a} \hat{\sigma}_- + \hat{a}^\dag \hat{\sigma}_+ \right),
\end{align}
where $\hat{a}$ is the annihilation operator of the photon field, $\hat \sigma_{x,y,z}$ are the Pauli matrices, $\hat{\sigma}_\pm =(\hat{\sigma}_x\pm i \hat{\sigma}_y)/2$, $\omega_c$ is the frequency of the bosonic mode, and $\Omega$ is the energy splitting of the two-level system  (we set $\hbar=1$). Here the interaction is written in terms of the couplings $\lambda_-$ and $\lambda_+$ corresponding to co-rotating and counter-rotating terms, respectively. Alternatively, the notation:
\begin{equation}\label{lambda_xy}
\lambda_{\pm} = \lambda_x \pm \lambda_y,
\end{equation}
is sometimes preferred, which emphasizes the coupling strength $\lambda_x$ ($\lambda_y$) of $\hat{J}_x$ ($\hat{J}_y$) to the $\hat{a}+\hat{a}^\dag$ ($i\hat{a}-i\hat{a}^\dag$) quadrature of the field  \cite{baksic2014controlling,Soriente2018}. The imbalance of the corotating and counter-rotating terms can be altered by the power of a bichromatic laser drive in trapped ion setups, or depends on the relative strength of the electric and magnetic couplings in circuit-QED implementations~\cite{wang2018quantum,puebla2017probing,cheng2018nonlinear,zhong2019quantum}. Considering an open environment causing photon decay, the model is described by the master equation:
\begin{align}
\label{eq-me}
\dot{\hat{\rho}}=-i[\hat{H},\hat{\rho}]+\kappa \mathcal{D}[\hat{a}]\hat{\rho},
\end{align}
where $\mathcal{D}[\hat{a}]\hat{\rho}=2\hat{a}\hat{\rho} \hat{a}^{\dagger}-\hat{a}^{\dagger}\hat{a}\hat{\rho}-\hat{\rho} \hat{a}^{\dagger}\hat{a}$. The above quantum evolution can be alternatively expressed as:
\begin{align}
\label{eq-motion}
&\frac{d}{dt} \hat{a}  =
-\left(\kappa+i\omega_c\right)\hat{a}+i\left(\lambda_- \hat{\sigma}_- +\lambda_+ \hat{\sigma}_+\right)+\hat{\mathcal{F}},
\notag \\
&\frac{d}{dt} \hat{\sigma}_+ =
i\Omega \hat{\sigma}_+ +i\left(\lambda_- \hat{a}^{\dagger} +\lambda_+ \hat{a} \right) \hat{\sigma}_z ,
\notag \\
&\frac{d}{dt} \hat{\sigma}_z  =
2i\left[\lambda_-\left( \hat{a}  \hat{\sigma}_+ - \hat{a}^{\dagger}  \hat{\sigma}_- \right)-\lambda_+\left( \hat{a} \hat{\sigma}_- - \hat{a}^{\dagger} \hat{\sigma}_+ \right)\right],
\end{align}
where the force $\hat{\mathcal{F}}(t)$ is a stochastic operator
satisfying $\langle \hat{\mathcal{F}}(t)\hat{\mathcal{F}}^{\dagger}(t')\rangle=2\kappa\delta(t-t')$ and $\langle {\hat{\mathcal{F}}}^{\dagger}(t')\hat{\mathcal{F}}(t)\rangle=0$.

In the rest of the paper, we will assume $\omega_c >0$ and, unless explicitly stated, $\Omega, \lambda_\pm >0$. To see that the latter restriction is not important, we first consider $\hat{U}_1=\exp[i\frac{\pi}{4}(\hat{\sigma}_z-2\hat{a}^\dag \hat{a})]$, which transforms the couplings as $\lambda_\pm \to \pm \lambda_\pm$. Instead, $\hat{U}_2=\exp[i\frac{\pi}{2} \hat{\sigma}_z]$ leads to a change of sign of both couplings, $\lambda_\pm \to - \lambda_\pm$. Through $U_{1}$ and $U_{2}$ we can always map the system to the region $\lambda_\pm >0$. Furthermore, $\hat{U}_3=\exp[i\frac{\pi}{2} \hat{\sigma}_x]$ changes the sign of the atomic level splitting, $\Omega \to -\Omega$, while interchanging the role of co-rotating and counter-rotating terms, $\lambda_\pm \to \lambda_\mp$. Therefore, we can also assume $\Omega>0$. We will also often use dimensionless couplings $\tilde\lambda_\pm = \lambda_\pm/\lambda_c$, where:
\begin{equation}\label{lc}
\lambda_c = \frac12 \sqrt{\omega_c \Omega}
\end{equation}
is the critical coupling to the superradiant phase in the closed ($\kappa =0 $) and isotropic ($\lambda_- = \lambda_+$) limit, both for the Dicke and Rabi model. However, while the usual quantum phase transition (QPT) of the Dicke model takes place for $N\to \infty$, for the Rabi model one should consider the following frequency ratio:
\begin{equation}\label{eta_def}
\eta =\frac{\Omega}{\omega_c}.
\end{equation}
and take the limit $\eta \to \infty$, which plays the same role of the conventional thermodynamic limit~\cite{levine2004entanglement,HinesPRA2004, BakemeierPRA2012, ashhab2013superradiance, hwang2015quantum, puebla2017probing, liu2017universal, hwang2018dissipative}.

\section{Mean-field analysis}\label{mean_field}

We fist examine the dissipative phase transition through standard mean-field theory, which is normally expected to become accurate in the thermodynamic limit $\eta\to \infty$. Here, however, since we are also interested in the stationary state of the system, the regime of validity should be carefully specified. As we will see, the problem arises due to the presence of multistability at arbitrary values of $\lambda_\pm$. In short, mean-field theory is valid when taking the limit $\eta \to \infty$ from the onset, i.e., before considering $t\to\infty$. This effectively suppresses the transitions between the stationary mean-field solutions, making them infinitely long-lived. A more precise discussion is given at the end of this section.  

From Eq.~(\ref{eq-motion}), mean-field equations for the expectation values $a=\langle \hat{a} \rangle$, $s_\pm = \langle \hat{\sigma}_\pm\rangle$, and $s_z = \langle \hat{\sigma}_z \rangle$ are obtained by neglecting quantum fluctuations, i.e., assuming the factorization $\langle \hat{A}\hat{B} \rangle=\langle \hat{A}\rangle\langle \hat{B}\rangle$. The system conserves the pseudo-angular momentum, thus we set $4|s_+|^2+s_z^2=1$. Here, as we are interested in the limit $\eta \to \infty$ instead of $N\to \infty$, we find that the following rescaling of the photon field and coupling strengths are particularly appropriate:
\begin{equation}
c = \frac{\langle \hat{a} \rangle}{\sqrt{\eta}},\qquad \tilde\lambda_\pm = \frac{\lambda_\pm}{\lambda_c},
\end{equation}
where $\eta$ is defined in Eq.~(\ref{eta_def}) and $\lambda_c $ is given in Eq.~(\ref{lc}). After also rescaling the time as $\tau= \omega_c t$, we find the mean-field equations:
\begin{align}
&\frac{dc}{d\tau } =-\left( \frac{\kappa }{\omega _{c}}+i\right) c+\frac{i}{%
2}\left( \tilde{\lambda}_- s_{-}+\tilde{\lambda}_+ s_{+}\right), \label{MFeqsc}\\
&\frac{1}{\eta}\frac{ds_+}{d\tau }=i \,  {\rm sgn}(\Omega) \, s_{+}+\frac{i}{2}
\left(\tilde{\lambda}_- c^{\ast }+\tilde{\lambda}_+ c\right) s_{z} , \label{MFeqssp} \\
&\frac{1}{\eta}\frac{d s_z}{d\tau } = i \tilde{\lambda}_-\left(
cs_{+}-c^{\ast }s_{-}\right) -i\tilde{\lambda}_+ \left( cs_{-}-c^{\ast }s_{+}\right). \label{MFeqssz} 
\end{align}%
The stationary states of these semiclassical equations are independent of $\eta$ and, actually, coincide with the stationary states of the Dicke model \cite{Soriente2018,Stitely2020}. As we will see, however, an important difference appears in the stability analysis. Since here we are interested in the limit $\eta \to \infty$, the left-hand-side of Eqs.~(\ref{MFeqssp}) and (\ref{MFeqssz}) vanish, which allows us to solve these equations explicitly. This reflects the fact that the spin follows adiabatically the instantaneous value of $c$ when $\Omega \gg \omega_c,\lambda_\pm$. Using the normalization condition $s_z^2 +4|s_+|^2 =1$ and assuming $\Omega>0$ we obtain:
\begin{align}
& s_z=\pm \frac{1}{\sqrt{1+|\tilde{\lambda}_-c^* + \tilde{\lambda}_+ c|^2 }}, \label{sz_MF} \\
& s_+ = -\frac12 (\tilde{\lambda}_-c^* + \tilde{\lambda}_+ c)s_z, \label{splus_MF}
\end{align}
where the $\pm$ sign is determined by the initial condition of $s_z$. In this limit, spin states in the upper (lower) half of the Bloch sphere never cross to the lower (upper) side. Finally, Eqs.~(\ref{MFeqsc}--\ref{MFeqssz}) can be written as an equation of motion for the cavity field alone:
\begin{equation}\label{ceq_MF}
\frac{dc}{d\tau } =-\left( \frac{\kappa }{\omega _{c}}+i\right) c\mp\frac{i}{4} \frac{(\tilde\lambda_-^2+\tilde\lambda_+^2)c+2\tilde\lambda_-\tilde\lambda_+c^*}{\sqrt{1+|\tilde{\lambda}_-c^* + \tilde{\lambda}_+ c|^2 }}.
\end{equation}

\begin{figure}
\centering
\includegraphics[width=0.4\textwidth]{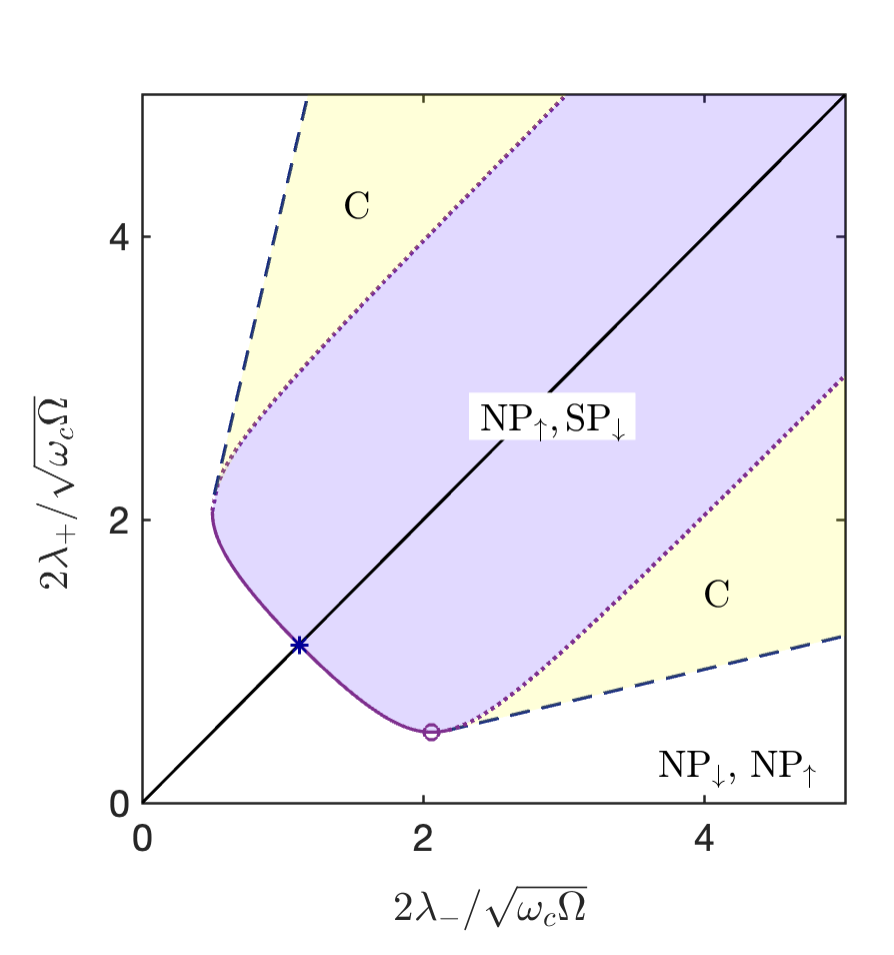}
\caption{\label{fig1-phase} Mean-field phase diagram at $\eta \to \infty$. Only stable stationary solutions are indicated. In the coexistence region C the three states (NP$_\uparrow$, NP$_\downarrow$, and SP$_\downarrow$) are all stable. The black line indicates the isotropic case ($\tilde\lambda_- = \tilde\lambda_+$), where the asterisk marks the critical point $\tilde\lambda_c^* = \sqrt{1+\kappa^2/\omega_c^2}$. The empty circle marks one of the tricritical points, with $\lambda_- = \tilde\lambda_c^{\rm T}$ [see Eq.~(\ref{lambdaT})]. Here we used $\kappa/\omega_c=0.5$. 
}
\end{figure}

Based on the mean-field equations, we obtain the phase diagram shown in Fig.~\ref{fig1-phase}. While at finite $\eta$ the long-time dynamics may be characterized by an oscillatory or chaotic behavior, resulting in a complex phase diagram \cite{Stitely2020}, the mean-field treatment becomes much simpler when $\eta\to \infty$. Here, we only need to consider the normal and superradiant fixed points. In the normal-phase (NP), the bosonic field has zero mean-field amplitude while the qubit can be either in  the  $|\downarrow\rangle$ state
\begin{align}
{\rm{NP}_{\downarrow}}: s_z=-1,\,\,\, s_+=0, \,\,\, c=0,
\end{align}
or fully populated in the up-state $|\uparrow\rangle$:
\begin{align}
{\rm{NP}_{\uparrow}}: s_z=1,\,\,\, s_+=0, \,\,\, c=0.
\end{align}
The stability analysis of the NP states can be performed by linearizing Eq.~(\ref{ceq_MF}) around $c=0$. It is then found that the NP$_\uparrow$ state is always stable, while NP$_\downarrow$ becomes unstable in the purple region of Fig.~\ref{fig1-phase}. It is convenient to specify the stability condition for a fixed ratio $r=\tilde\lambda_+ /\tilde\lambda_-$. Then, NP$_\downarrow$ is unstable for
\begin{equation}
\tilde\lambda_c^- < \tilde\lambda_- < \tilde\lambda_c^+,
\end{equation} 
where we defined:
\begin{equation}\label{lambdac_pm}
\tilde\lambda_c^\pm = \frac{2\sqrt{1+r^2\pm \sqrt{4r^2-(1-r^2)^2\kappa^2/\omega_c^2}}}{|1-r^2|}.
\end{equation}
In Fig.~\ref{fig1-phase}, the solid phase boundary of the purple region is at $\tilde\lambda_-=\tilde\lambda_c^-$ (while $\tilde\lambda_+=r \tilde\lambda_c^-$) and the two dotted boundaries are at $\tilde\lambda_-=\tilde\lambda_c^+$. Note that in Eq.~(\ref{lambdac_pm}) $r$ should satisfy $r_- \leq  r \leq r_+$, where:
\begin{equation}\label{r_pm}
r_\pm = \sqrt{1+\omega_c^2/\kappa^2} \pm \omega_c/\kappa.
\end{equation}
When $r <r_-$ or $r>r_+$, both NP$_\downarrow$ and NP$_\uparrow$ are stable fixed points for arbitrary values of $\tilde\lambda_-$.

Besides NP states, the following superradiant fixed-point plays an important role:
\begin{align}
\label{eq-sp}
{\rm{SP}}_\downarrow :& \,\, s_z=-\cos{\theta_{\rm SP}}, \quad s_+=\pm\frac{1}{2}e^{i\phi_{\rm SP}}\sin\theta_{\rm SP}\nonumber \\
& c =\frac{ \tilde{\lambda}_- s_-+\tilde{\lambda}_+ s_+}{2(1-i\kappa/\omega_c)},
\end{align}
where the direction of the spin is given by:
\begin{align}\label{SP_spin_direction}
&\cos{\theta_{\rm SP}} = 
\frac{4\left(1+\kappa^2/\omega_c^2\right)}
{\tilde{\lambda}_-^2+\tilde{\lambda}_+^2+
\sqrt{4\tilde{\lambda}_+^2\tilde{\lambda}_-^2-\frac{\kappa^2}{\omega^2_c}\left(\tilde{\lambda}_-^2-\tilde{\lambda}_+^2\right)^2}},  \nonumber \\
&\sin{2\phi_{\rm SP}}=-\frac{\kappa}{\omega_c}\frac{\tilde{\lambda}_-^2-\lambda_+^2}{2\tilde{\lambda}_-\tilde{\lambda}_+} .
\end{align}
The stability analysis is rather involved and we must resort to numerical linearization around the steady-state. The final result is that SP$_\downarrow$ exists in the purple and yellow regions of Fig.~\ref{fig1-phase}, where it is always stable. The phase boundary can be divided into a second-order critical line (marked as a solid curve in Fig.~\ref{fig1-phase}) and first-order transitions (dashed lines). The second-order critical line can be parameterized by $r$ as $\tilde\lambda_-  = \tilde\lambda_c^-$ and $\tilde\lambda_+ = r \tilde\lambda_c^- $, where $\tilde\lambda_c^-$ is given in Eq.~(\ref{lambdac_pm}). This curve ends at $r=r_\pm$ with two tricritical points \cite{Soriente2018,Stitely2020}. The one marked by an open circle in Fig.~\ref{fig1-phase} is at $\tilde\lambda_- = \tilde\lambda_c^{\rm T}$ and $\tilde\lambda_+ = r_-\tilde\lambda_c^{\rm T}$, where $r_-$ is given in Eq.~(\ref{r_pm}) and:
\begin{equation}\label{lambdaT}
\tilde\lambda_c^{\rm T} =\sqrt{2\left(1+\kappa^2/\omega_c^2 + \sqrt{1+\kappa^2/\omega_c^2}\right)}.
\end{equation} 

Beyond the tricritical points, the first-order phase boundaries (dashed lines in Fig.~\ref{fig1-phase}) are simply given by $\tilde\lambda_{+} = r_\pm \tilde\lambda_-$. Within the mean-field approximation, when entering the normal phase across these boundaries the order parameter $|c|^2$ of the superradiant state drops discontinuously to zero. We also mention that another superradiant solution exists in the yellow regions of Fig.~\ref{fig1-phase}. However, this fixed point is always unstable and does not appear in the phase diagram. Furthermore, as stated already, Eq.~(\ref{ceq_MF}) does not give any chaotic or oscillatory phase.

In closing this section about the mean-field analysis, we expand upon our earlier discussion about the regime of applicability. While mean-field theory becomes accurate in the limit $\eta \to \infty$, we are interested here in the stationary state of the system, i.e., we also impose $t \to \infty$. It is then crucial to specify in which order the two limits are applied. From the derivation of Eq.~(\ref{ceq_MF}), it is clear that the phase diagram of Fig.~\ref{fig1-phase} is obtained by taking $\eta \to \infty$ before $t\to \infty$. More precisely, as we will establish explicitly later on, the mean-field treatment is only valid on a timescale $t_{\rm mf} $ which is proportional to $\eta$:
\begin{equation}\label{t_mf}
t \ll t_{\rm mf} \propto \eta  . 
\end{equation}
If we take $\eta \to \infty$ from the onset, the above restriction does not play any role. The mean-field theory predicts correctly that, in this limit, the states appearing in Fig.~\ref{fig1-phase} are stable for an arbitrarily long time.

\begin{figure}
\centering
\includegraphics[width=0.4\textwidth]{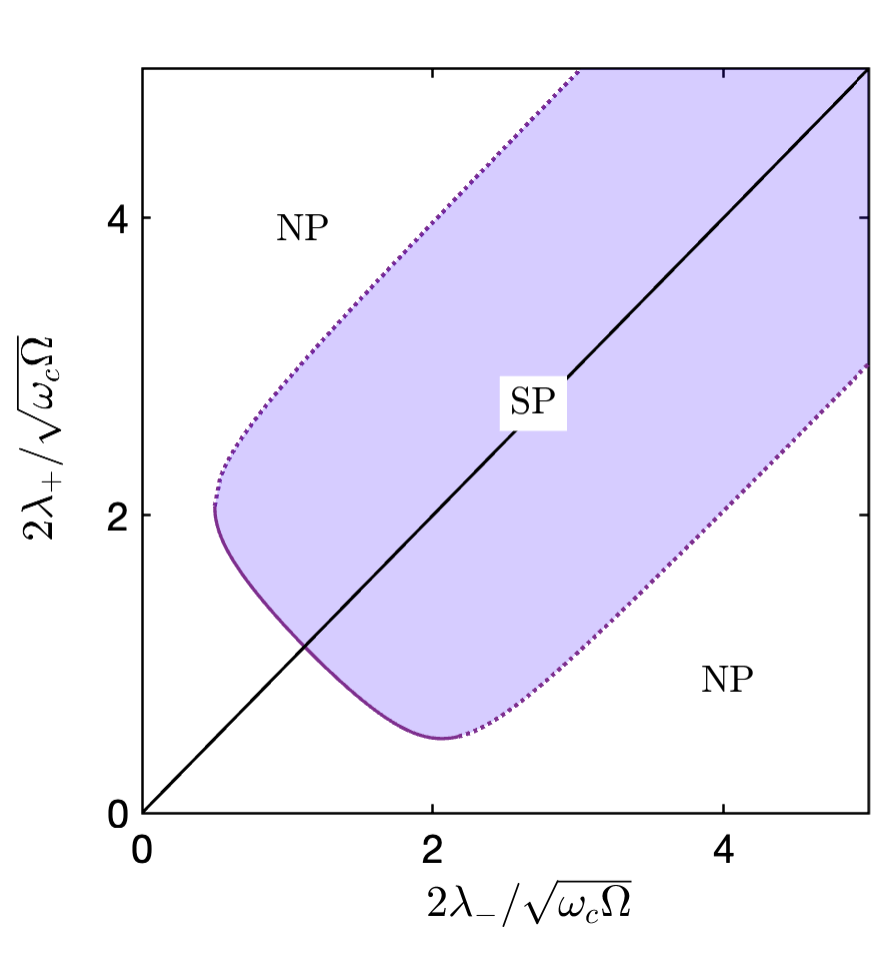}
\caption{\label{fig2-phase} Mean-field phase diagram obtained by considering the unique steady state of Eq.~(\ref{eq-me}) in the limit $\eta \to \infty$.  In the NP phase the system is in a statistical mixture of NP$_\uparrow$ and NP$_\downarrow$, while the SP phase is formed by statistical mixtures of NP$_\uparrow$ and SP$_\downarrow$. The dotted and solid curves mark first- and second-order critical lines, respectively. We used the same parameters of Fig.~\ref{fig2-phase}. 
}
\end{figure}

Alternatively, one might be interested in the properties of the steady state at finite but arbitrarily large $\eta$. This is a different question which assumes the opposite order of limits, i.e., $t\to \infty$ before $\eta \to \infty$. Since the condition Eq.~(\ref{t_mf}) is violated, the phase diagram of Fig.~\ref{fig1-phase} gets modified. In particular, multi-stable phases are not allowed in this limit, since the quantum master equation has always a unique steady-state. The phase diagram of the steady state at large $\eta$ (i.e., considering the regime $t \gg t_{\rm mf}$)  is shown in Fig.~\ref{fig2-phase}, which summarizes some of the main conclusions of the following sections. In this phase diagram there are only a normal (NP) and superradiant (SP) phase. In each phase, the system is in a well-defined statistical mixture of the mean-field states: NP$_{\downarrow}$ and NP$_{\uparrow}$ (SP$_\downarrow$ and NP$_{\uparrow}$) in the former (latter) case. Notably, the C phase of Fig.~\ref{fig1-phase} has disappeared, and the dotted line has become a first-order phase boundary between the normal and superradiant phases.

\section{Finite-$\eta$ effects}\label{sec:finite_eta}

We now start our analysis of quantum fluctuations, which determine the fate of the general multi-stable behavior found in Fig.~\ref{fig1-phase}. We first consider the simpler yet instructive treatment of the normal phase. The cumulant expansion is a rigorous approach to include quantum effects, and we show that it indeed allows us to derive in a systematic manner the correct stationary properties of the system (Sec.~\ref{cumulant}). However, the cumulant expansion is not physically very transparent, and it is cumbersome to apply it to the superradiant state. This is why we switch to an effective master-equation approach, recovering the correct stationary properties of the normal phase (Sec.~\ref{effective-me}). We then generalize this treatment to the superradiant phase (Sec.~\ref{sec:second_order_transition}) and to the tristable phase (Sec.~\ref{sec:Cphase}), to establish the complete phase diagram shown in Fig.~\ref{fig2-phase}.

\subsection{Cumulant expansion in the normal phase} \label{cumulant}  

In mean-field theory, we have used the factorization assumption $\langle \hat{A}\hat{B} \rangle=\langle \hat{A}\rangle\langle \hat{B}\rangle$ to decouple qubit-bosonic field correlations and obtained a set of dynamical equations which only depend on $s_z,s_+$ and the (properly rescaled) bosonic field amplitude, $\langle \hat{a} \rangle$. Beyond the mean-field approach, qubit-bosonic field correlations should be taken into account. For example, from Eq.~(\ref{eq-motion}) we immediately obtain:
\begin{equation}
\label{eq-sz_exact}
\frac{d \langle\hat{\sigma}_z \rangle}{dt} 
=2i[ \lambda_-(\langle\hat{a}\hat{\sigma}_+ \rangle -\langle \hat{a}^{\dagger}\hat{\sigma}_- \rangle) 
-\lambda_+(\langle \hat{a}\hat{\sigma}_-\rangle -\langle\hat{a}^{\dagger}\hat{\sigma}_+ \rangle) ],
\end{equation}
where the evolution of the second-order correlators can be described through the cumulant expansion~\cite{oztop2012excitations,kirton2018superradiant}. To leading-order, we assume that third-order moments are zero, i.e., set $\langle \hat{A}\hat{B}\hat{C}\rangle=\langle \hat{A}\hat{B}\rangle\langle \hat{C}\rangle+\langle \hat{A}\hat{C}\rangle\langle \hat{B}\rangle+\langle \hat{B}\hat{C}\rangle\langle \hat{A}\rangle-2\langle \hat{A}\rangle\langle \hat{B}\rangle\langle \hat{C}\rangle$. By doing so, we obtain a set of non-linear equations such as:
\begin{align}
\label{eq-cumul-motion}
\frac{d\langle \hat{a}\hat{\sigma}_+ \rangle}{dt}
=&-(\kappa+i\omega_c-i\Omega)\langle\hat{a}\hat{\sigma}_+ \rangle 
+\frac{i\lambda_-}{2}\left(1+\langle \hat{\sigma}_z \rangle\right)\nonumber \\
& + i (\lambda_-\langle\hat{a}^{\dagger}\hat{a}\hat{\sigma}_z\rangle
+ \lambda_+ \langle\hat{a}^2\hat{\sigma}_z \rangle ),
\end{align}
where the two last terms are approximated by:
\begin{align}
\langle \hat{a}^\dag\hat{a} \hat{\sigma}_z \rangle = &\langle \hat{a}^\dag\hat{a} \rangle\langle\hat{\sigma}_z \rangle+  \langle \hat{a}^\dag\hat{\sigma}_z \rangle \langle \hat{a}  \rangle+ \langle \hat{a}\hat{\sigma}_z \rangle \langle \hat{a}^\dag  \rangle \nonumber\\
&- 2\langle \hat{a} \rangle\langle \hat{a}^\dag \rangle \langle \hat{\sigma}_z \rangle, \nonumber \\
\langle \hat{a}^2 \hat{\sigma}_z \rangle = & \langle \hat{a}^2 \rangle\langle\hat{\sigma}_z \rangle+ 2 \langle \hat{a}\hat{\sigma}_z \rangle \langle \hat{a}  \rangle- 2\langle \hat{a} \rangle^2 \langle \hat{\sigma}_z \rangle.
\end{align}
The complete set of equations for the second-order correlators is presented in Appendix~\ref{app-cumulant}. 

In the normal phase (i.e., the white region of Fig.~\ref{fig1-phase}) the order parameter $\langle \hat{a} \rangle$ is zero, which allows us to simplify the non-linear equations and obtain analytic expressions for the stationary expectation values. The qubit population is given by:
\begin{align}
\label{eq-cumul}
\langle \hat{\sigma}_z \rangle \simeq \frac{\lambda_+^2-\lambda_-^2}{\lambda_-^2+\lambda_+^2}+\Delta,
\end{align}
where $\Delta$ is a small ($\sim  1/\eta$) correction, whose explicit expression is given in Appendix~\ref{app-cumulant}. Higher-order terms are omitted, as they are not significant (to derive them, it would be necessary to include higher-order correlators in the cumulant expansion). 

The leading term of Eq.~(\ref{eq-cumul}) already captures the effects of quantum fluctuations, which are beyond mean-field theory. In the normal phase, mean-field theory predicts that both NP$_\uparrow$ and NP$_\downarrow$ are stable fixed points with distinct basins of attraction, thus the stationary expectation value at $t \to \infty$ is $s_z = \pm 1$, depending on the initial condition. Instead, Eq.~(\ref{eq-cumul}) gives a very different result. For example, in the balanced case $\lambda_- = \lambda_+$ we obtain a small expectation value $\langle \hat{\sigma}_z \rangle \simeq -2/\eta$. This difference is due to fluctuations between the two mean-field solutions, induced by the quantum evolution. At any finite $\eta \gg 1$, the exact dynamics described by Eq.~(\ref{eq-me}) leads to a stationary state which is a well-defined statistical mixture of $\rm{NP}_{\uparrow}$ and $\rm{NP}_{\downarrow}$. From Eq.~(\ref{eq-cumul}) we can infer that, to leading order in $\eta$, the populations of these two states are respectively given by $P_+$ and $P_-$,  where:
\begin{equation}\label{Ppm}
P_\pm = \frac{\lambda_\pm^2}{\lambda_+^2+\lambda_-^2}.\quad {\rm (normal~phase)}
\end{equation}

\begin{figure}
\centering
\includegraphics[width=0.45\textwidth]{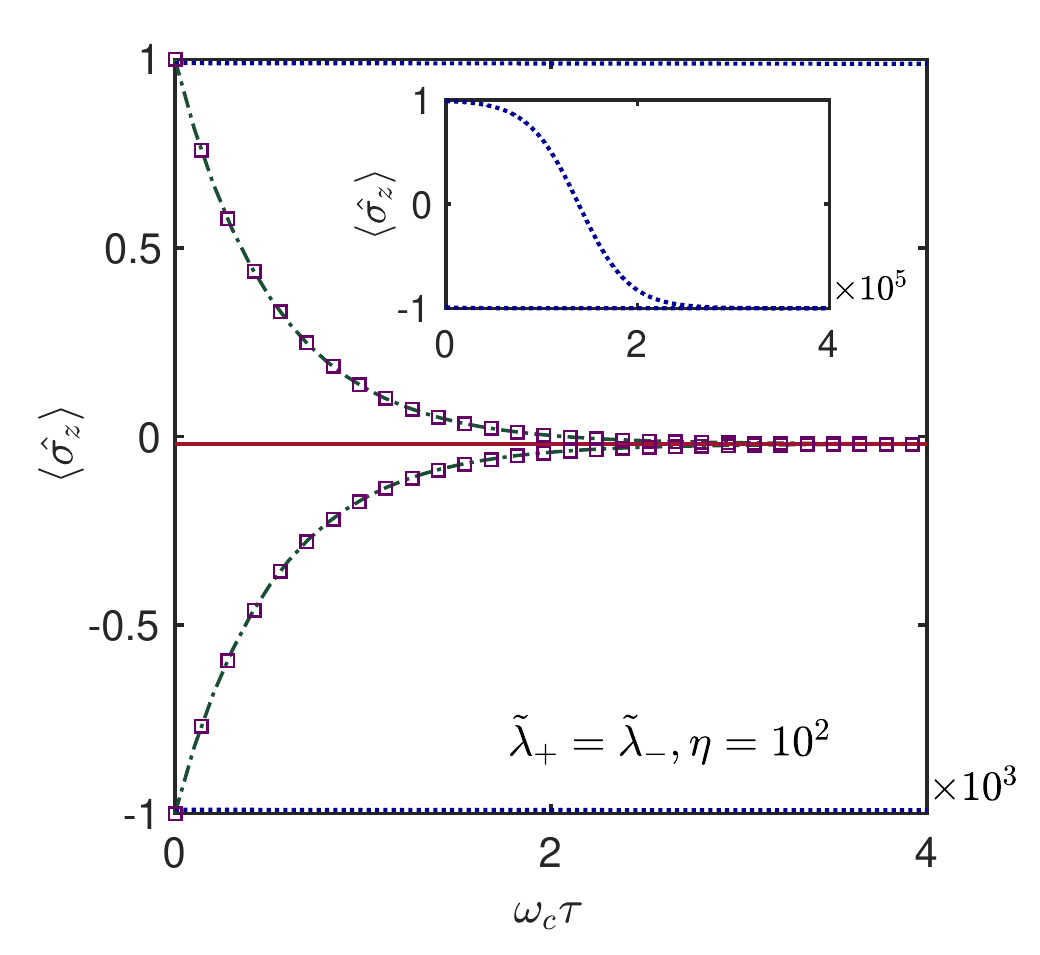}
\caption{\label{fig3} Time evolution of $\langle \hat{\sigma}_z \rangle$ for the isotropic Rabi model ($\tilde\lambda_+=\tilde\lambda_-$) at a finite value $\eta= 100$. The purple squares are calculated with the master equation (\ref{eq-me}), considering both $|\uparrow\rangle$ and $|\downarrow\rangle$ as initial states (the bosonic mode is initially in the vacuum). The green dashed-dotted lines are computed numerically from the cumulant expansion, see Eq.~(\ref{eq-cumul-motion}) and Appendix~\ref{app-cumulant}, and show excellent agreement with the exact dynamics. The red line is the approximate stationary value of Eq.~(\ref{eq-cumul}). The blue dotted lines are calculated based on the mean-field equations~(\ref{MFeqsc}--\ref{MFeqssz}). In the main panel, the mean-field evolution does not display any significant time dependence. For completeness, we also show the mean-field evolution for much longer times (inset). We used the parameters $\kappa=0.5\omega_c$ and  $\tilde\lambda_+=\tilde\lambda_-=0.618$.
}
\end{figure}

To verify the validity of the cumulant expansion, we have performed an exact calculation based on Eq.~(\ref{eq-me}), see the purple squares in Fig.~\ref{fig3}. In this numerical example we consider the isotropic model ($\lambda_-=\lambda_+$) and two initial conditions, corresponding to either NP$_\uparrow$ or NP$_\downarrow$. As seen from Fig.~\ref{fig3}, mean-field theory (dotted curves) predicts that the two initial values $\langle \hat{\sigma}_z \rangle =  \pm 1$ are constant on a timescale which is much longer than the true relaxation time. Instead, the numerical solutions obtained from the cumulant expansion (green dashed-dotted curves) are in excellent agreement with the master equations. The stationary value of $\langle \hat{\sigma}_z \rangle$ is also very close to the one predicted by Eq.~(\ref{eq-cumul}).

Figure~\ref{fig3} allows us to discuss more concretely the previous arguments about the $\eta \to \infty$ and $t\to \infty$ limits. The semiclassical treatment becomes formally exact when $\eta \to \infty$, but Fig.~\ref{fig3} shows that this is only true if the evolution is shorter than the typical relaxation time associated with the quantum evolution. As seen, mean-field theory fails over the long evolution time considered in Fig.~\ref{fig3} (where $\tau \gtrsim 10^3 \gg \eta$). If in Fig.~\ref{fig3} we restrict ourselves to shorter times (e.g., $\tau \ll \eta$), the discrepancy between exact evolution and semiclassical dynamics remains small. It is also important to note that the relaxation time of the exact quantum dynamics grows with $\eta$, so mean-field theory becomes accurate for longer time periods when taking the $\eta \to \infty$ limit.

The cumulant expansion can be also applied to the other multi-stable phases of Fig.~\ref{fig1-phase}. However, in this case we cannot easily obtain analytic results. For this reason, we prefer the alternative approach described in the following sections. Working to leading order in $\eta$, we will derive effective master equations which describe in a transparent manner the transitions between different fixed points. In the normal phase, we can recover the first term of Eq.~(\ref{eq-cumul}) and the treatment can be extended to the rest of the phase diagram.

\subsection{Effective master equation in the normal phase}
\label{effective-me}

The derivation of the effective master equation in the normal phase is relatively straightforward. As a first step, we apply a unitary transformation $\hat{U}_{np}$, which approximately decouples the qubit and cavity field \cite{hwang2015quantum,liu2017universal}:
\begin{align}
\label{Unp}
\hat{U}_{np}=e^{\frac{1}{2\sqrt{\eta}}\left[\left(\tilde{\lambda}_-\hat{a}+\tilde{\lambda}_+ \hat{a}^{\dagger}\right)\hat{\sigma}_{+}-\left(\tilde{\lambda}_+ \hat{a}+\tilde{\lambda}_-\hat{a}^{\dagger}\right)\hat{\sigma}_{-}\right]} .
\end{align}
By neglecting terms of order $\omega_c/\sqrt{\eta}$ in the effective Hamiltonian, $H_{np}=\hat{U}_{np}^{\dagger}\hat{H}\hat{U}_{np}$, we obtain:
\begin{align}
\label{Hnp}
\hat{H}_{np}&\simeq \,  
 \frac{\Omega}{2}\left( 1+\frac{\tilde{\lambda}_-^2+\tilde{\lambda}_+^2}{4\eta}\right)\hat{\sigma}_z + \omega_c \hat{a}^{\dagger}\hat{a} \nonumber\\
& +\frac{\omega_c}{4}
\left[\left(\tilde{\lambda}_-^2+\tilde{\lambda}_+^2\right) \hat{a}^{\dagger}\hat{a}
+\tilde{\lambda}_- \tilde{\lambda}_+\left(\hat{a}^2+\hat{a}^{\dagger2}\right)\right]\hat{\sigma}_z .
\end{align}
Since $\hat{H}_{np}$ is diagonal in $\hat\sigma_z$ and, to leading order in $\eta$, only cavity decay takes place, the $\hat\sigma_z = \pm 1$ subspaces are decoupled also for the dissipative time evolution. This leading-order approximation is consistent with the mean-field treatment, when there are two stable fixed points NP$_\uparrow$ and NP$_\downarrow$ with no transition taking place between them. However, to describe the evolution of $\hat{\widetilde{\rho}} = \hat{U}_{np}^{\dagger}\hat{\rho}\hat{U}_{np}$ more accurately, it is necessary to consider how the Lindblad term is affected by the unitary transformation. The hybridization with the cavity field has the well-known consequence of inducing decay of the two-level atom (i.e., Purcell relaxation). For completeness, the transformation of the Lindlad term is discussed in detail in Appendix~\ref{app-e-me}, while we include below the most important effect:
\begin{align}
\label{eq-effectiveME-normal}
\dot{\hat{\widetilde{\rho}}}=& -i\left[\hat{H}_{np},\hat{\widetilde{\rho}}\right]+\kappa\mathcal{D}[\hat{a}]\hat{\widetilde{\rho}} +\frac{\kappa}{4\eta}\sum_{\mu = \pm} \tilde{\lambda}^2_\mu \mathcal{D}[\hat{\sigma}_\mu]\hat{\widetilde{\rho}}.
\end{align}
The last term of Eq.~(\ref{eq-effectiveME-normal}) describes spin flips beyond the mean-field approximation. As mentioned in Eq.~(\ref{t_mf}), these spin-flips occur on a long timescale $\sim \eta/\kappa$, but are essential to establish the unique stationary state at any finite $\eta$.   Since the interaction contains co- and counter-rotating terms, both relaxation and excitation processes appear. The corresponding rates:
\begin{align}
\Gamma_{{\rm NP}_\uparrow \to {\rm NP}_\downarrow} = \frac{\tilde{\lambda}_-^2}{2\eta} \kappa, \label{Gamma_NP1}\\
\Gamma_{{\rm NP}_\downarrow \to {\rm NP}_\uparrow} = \frac{\tilde{\lambda}_+^2}{2\eta} \kappa, \label{Gamma_NP2}
\end{align} 
give the populations in Eq.~(\ref{Ppm}) and the leading term of Eq.~(\ref{eq-cumul}). By interpreting the rates in terms of an effective temperature $T_{\rm eff}$, we immediately have (with $k_{\rm B}=1$):
\begin{equation}
T_{\rm eff} = \frac{\Omega}{2\ln (\lambda_-/\lambda_+)}.
\end{equation}
Interestingly, the incoherent pumping induced by the counter-rotating terms leads to an infinite-temperature spin state in the isotropic model ($\lambda_+ = \lambda_-$), while population inversion is realized when $\lambda_+ > \lambda_-$.

\subsection{Second-order NP-SP transition}\label{sec:second_order_transition}

After having discussed the normal phase, we now consider the superradiant states. As we will see, the discussion of transition rates becomes more subtle than in the normal phase. In this subsection we characterize the stationary state in the purple region of Fig.~\ref{fig1-phase}, where the stable fixed points of mean-field theory are NP$_\uparrow$ and SP$_\downarrow$. The coexistence region C, where also NP$_\downarrow$ is stable, will be discussed in the next subsection. 

To derive an effective master equation which is valid close to the superradiant fixed points, we should take into account the finite expectation value of the cavity field. We consider one of the two mean-field solutions for $c$, while the other broken-symmetry state can be treated in a completely analogous manner. It is convenient to first perform a displacement transformation $\hat{U}_{d}=e^{\sqrt{\eta} \left(c\hat{a}^{\dagger}-c^*\hat{a}\right)}$ \cite{hwang2015quantum}, giving the effective Hamiltonian $\hat{H}_{d} = \hat{U}_{d}^\dag\hat{H}\hat{U}_{d} +i \sqrt{\eta}\kappa(c^* \hat{a}-c\hat{a}^\dag)$. Furthermore, since the natural spin quantization axis is tilted away from the $z$ direction, we define:
\begin{align}
\label{eq-tau-def}
\vec{\hat{\tau}}  
=Y(\theta_{\rm SP})Z(-\phi_{\rm SP}) \vec{\hat{\sigma}} ,
\end{align}
where $Z(\pm\phi_{\rm SP})$ and $Y(\theta_{\rm SP})$ are rotation matrices around the $z$ and $y$ axis, respectively, with $\theta_{\rm SP},\phi_{\rm SP}$ given by Eq.~(\ref{SP_spin_direction}). After these preliminary transformations, we obtain:
\begin{align}
\label{Hd}
\hat{H}_{d}= \, &  \omega_c\hat{a}^{\dagger}\hat{a}+\frac{\Omega}{2\cos\theta_{\rm SP}}\hat{\tau}_z
-\frac{\sqrt{\eta} \omega_c}{2}(1+\hat{\tau}_z) \left(\tilde{\chi}_z\hat{a}+\tilde{\chi}^*_z\hat{a}^\dag\right)\notag \\
& ~ -\frac{\sqrt{\eta} \omega_c}{2} \left[\tilde{\chi}_-\hat{\tau}_+\hat{a}+\tilde{\chi}_+\hat{\tau}_+\hat{a}^{\dagger}+ {\rm H. c.}\right],
\end{align}
where the dimensionless couplings are:
\begin{align}
& \tilde\chi_\pm =\tilde\lambda_ \pm e^{i\phi_{\rm SP}}\cos^2\frac{\theta_{\rm SP}}{2} + \tilde\lambda_\mp e^{-i\phi_{\rm SP}} \sin^2\frac{\theta_{\rm SP}}{2} , \nonumber \\
& \tilde\chi_z= -\frac12 \sin\theta_{\rm SP}\left(\tilde\lambda_- e^{i\phi_{\rm SP}}+\tilde\lambda_+ e^{-i\phi_{\rm SP}}\right).
\end{align}

The second line of Eq.~(\ref{Hd}) contains co- and counter-rotating interactions with modified strength, given by $\tilde \chi_\pm$. To leading order, we can eliminate these off-diagonal terms by applying a unitary transformation $\hat{U}_{sp}$ analogous to Eq.~(\ref{Unp}). Details are given in Appendix~\ref{app-e-me}, where we derive the following master equation:
\begin{align}
\label{eq-effectiveME-sp}
\dot{\hat{\tilde{\rho}}}=&-i\left[\hat{H}_{sp},\hat{\bar{\rho}}\right]+\kappa\mathcal{D}[\hat{a}]\hat{\bar{\rho}} \notag \\
&+\frac{\kappa }{4\eta}\cos^2{\theta_{\rm SP}}\left(|\tilde\chi_-|^2\mathcal{D}[\hat{\tau}_-]\hat{\bar{\rho}}+|\tilde\chi_+|^2\mathcal{D}[\hat{\tau}_+]\hat{\bar{\rho}}\right). 
\end{align}
In the above Eq.~(\ref{eq-effectiveME-sp}), the effective Hamiltonian reads:
\begin{align}
\label{Hsp}
& \hat{H}_{sp} \simeq\frac{\Omega}{2}\left( \frac{1}{\cos\theta_{\rm SP}} +\frac{|\tilde\chi_-|^2+|\tilde\chi_+|^2}{4\eta}\cos\theta_{\rm SP}\right)\hat{\tau}_z + \omega_c\hat{a}^{\dagger}\hat{a} \nonumber \\
& +\frac{\omega_c}{4}\left[(|\tilde\chi_-|^2+|\tilde\chi_+|^2)\hat{a}^{\dagger}\hat{a} 
+(\tilde\chi_-\tilde\chi_+^*\hat{a}^2+\tilde\chi_-^*\tilde\chi_+\hat{a}^{\dagger2})\right]\hat{\tau}_z, \nonumber \\
& -\frac{\sqrt{\eta}\omega_c}{2}(\tilde\chi_z \hat{a}+ \tilde\chi_z^* \hat{a}^{\dag} )(1+\hat{\tau}_z).
\end{align}
Despite the more complex transformation (here the density matrix is $\hat{\bar{\rho}}=\hat{U}_{sp}^{\dagger} \hat{U}_d^{\dagger} \hat{\rho} \hat{U}_d \hat{U}_{sp}$), the type of spin-flip processes appearing in the second line of Eq.~(\ref{eq-effectiveME-sp}) are completely analogous to the last term of Eq.~(\ref{eq-effectiveME-normal}). In the superradiant case, $\tilde\chi_\pm$ and $\hat\tau_\alpha$ play the role of $\tilde\lambda_\pm$ and $\hat\sigma_\alpha$, respectively. The spin-flip rates are now given by:
\begin{align}
\Gamma_{{\rm SP}_\uparrow \to {\rm SP}_\downarrow} = \frac{|\tilde{\chi}_-|^2}{2\eta} \kappa \cos^2\theta_{\rm SP}, \label{SP_rate1} \\
\Gamma_{{\rm SP}_\downarrow \to {\rm SP}_\uparrow} = \frac{|\tilde{\chi}_+|^2}{2\eta} \kappa \cos^2\theta_{\rm SP}, \label{SP_rate2}
\end{align} 
where the additional factors $\cos^2 \theta_{\rm SP}$ are related to the fact that the $\tau_z = \pm 1$ eigenstates have spin polarization $\langle \hat\sigma_z \rangle = \pm \cos\theta_{\rm SP}$. On the other hand, if we consider the Hamiltonian in Eq.~(\ref{Hsp}), we find an important difference from the normal phase. The first three lines have the same structure of the normal-phase Hamiltonian $\hat{H}_{np}$, see Eq.~(\ref{Hnp}), but the last term of $H_{sp}$ has no analogy to the treatment of the normal state and leads to important physical consequences described below.

The third line of Eq.~(\ref{Hsp}) is a spin-dependent drive of the cavity, which vanishes when $\hat{\tau}_z = -1$. Therefore, if we restrict ourselves to the subspace $\hat{\tau}_z = -1$, the drive term is absent and the system will relax to a stationary state around $\langle \hat{a}^\dag \hat{a}\rangle =0$ (in the displaced frame). Instead, in the subspace $\hat{\tau}_z = 1$ the drive term is large and leads to a rapid increase of the average photon number, bringing the system beyond the regime of applicability of Eqs.~(\ref{eq-effectiveME-sp}) and (\ref{Hsp}). The interpretation of this behavior is straightforward from the point of view of the mean-field treatment: SP$_\downarrow$ is a stable fixed point, thus the evolution with $\hat{\tau}_z = -1$ leads to a well-defined stationary state. However, the state SP$_\uparrow$ (obtained from SP$_\downarrow$ by flipping the spin direction) is not a stationary mean-field solution. SP$_\uparrow$ undergoes a fast evolution which brings the photon field away from the expectation value $\langle \hat a \rangle \simeq c\sqrt{\eta}$ (in the original frame), and eventually leads to the other stable fixed point, NP$_\uparrow$.

\begin{figure}
\centering
\includegraphics[width=0.45\textwidth]{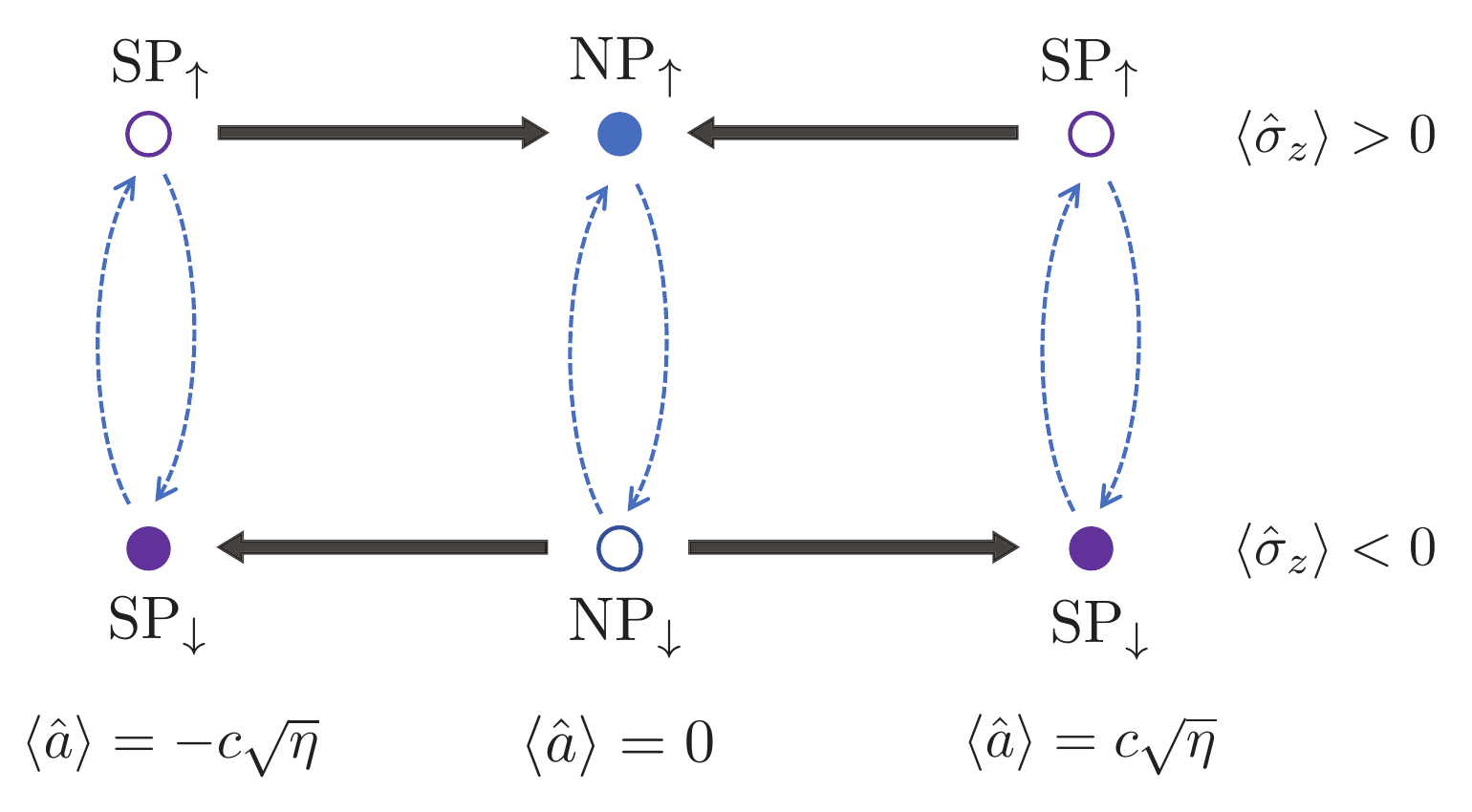}
\caption{\label{schematics_rates_1} Relevant transitions taking place in the superradiant phase. The filled dots (NP$_\uparrow$ and SP$_\downarrow$) are stable fixed points within the mean-field approximation, while the empty dots (NP$_\downarrow$ and SP$_\uparrow$) are unstable states. Fast transitions are indicated by thick arrows. The gray dashed arrows indicate slow spin flips described by the effective master equations. The corresponding rates (all $\propto \kappa/\eta$) are given in Eqs.~(\ref{Gamma_NP1}), (\ref{Gamma_NP2}), (\ref{SP_rate1}), and (\ref{SP_rate2}). }
\end{figure}


We now take into account all slow spin-flip processes, by including both normal and superradiant states. For convenience, the various transitions discussed are schematically represented in Fig.~\ref{schematics_rates_1}. First, we consider SP$_\downarrow$ as the initial state. On a long time scale corresponding to Eq.~(\ref{SP_rate2}) a spin flip takes place, bringing the system to SP$_\uparrow$. Afterward, SP$_\uparrow$ evolves to NP$_\uparrow$ within a short timescale. As described in the previous paragraph, this SP$_\uparrow \to$~NP$_\uparrow$ process is governed by the mean-field evolution, thus the corresponding transition rate is not suppressed by $\eta$. Since ${\rm SP}_\uparrow \to {\rm NP}_\uparrow$ can be considered as essentially instantaneous, the small spin-flip rate $\Gamma_{{\rm SP}_\uparrow \to {\rm SP}_\downarrow}$ in Eq.~(\ref{SP_rate1}) does not play any significant role and we obtain a dominant chain of transitions SP$_\downarrow \to $~SP$_\uparrow \to$~NP$_\uparrow$, with the effective rate given by Eq.~(\ref{SP_rate2}):
\begin{equation}\label{eff_rate_SR1}
\Gamma_{{\rm SP}_\downarrow \to {\rm NP}_\uparrow} \simeq \frac{|\tilde{\chi}_+|^2}{2\eta} \kappa \cos^2\theta_{\rm SP}. 
\end{equation}
Similarly, if the initial state is NP$_\uparrow$  a slow spin-flip process to NP$_\downarrow$ takes place, with a rate given by Eq.~(\ref{Gamma_NP1}). Since NP$_\downarrow$ is unstable in the superradiant phase, the system quickly reaches ${\rm SP}_\downarrow$. The effective rate for the whole process, ${\rm NP}_\uparrow \to{\rm NP}_\downarrow \to {\rm SP}_\downarrow$,  is given by:
\begin{equation}\label{eff_rate_SR2}
\Gamma_{{\rm NP}_\uparrow \to {\rm SP}_\downarrow} \simeq \frac{\tilde{\lambda}_-^2}{2\eta} \kappa . 
\end{equation}
From Eqs.~(\ref{eff_rate_SR1}) and (\ref{eff_rate_SR2}) we find the probability of being in the superradiant state:
\begin{equation}\label{Pm_SR}
P_{{\rm SP}_\downarrow} = \frac{\tilde\lambda_-^2}{\tilde\lambda_-^2+|\tilde\chi_+|^2\cos^2\theta_{\rm SP}},
\end{equation}
while the population of NP$_\uparrow$ is simply given by $P_{{\rm NP}_\uparrow} = 1- P_{{\rm SP}_\downarrow}$. At the second-order phase boundary (i.e., the solid curve in Fig.~\ref{fig1-phase}), these populations join smoothly with the $P_\pm$ of Eq.~(\ref{Ppm}), since $\theta_{\rm SP} \to 0$ and $|\tilde\chi_+|^2 \to \tilde\lambda_+^2$.

\begin{figure}
\centering
\includegraphics[width=0.48\textwidth]{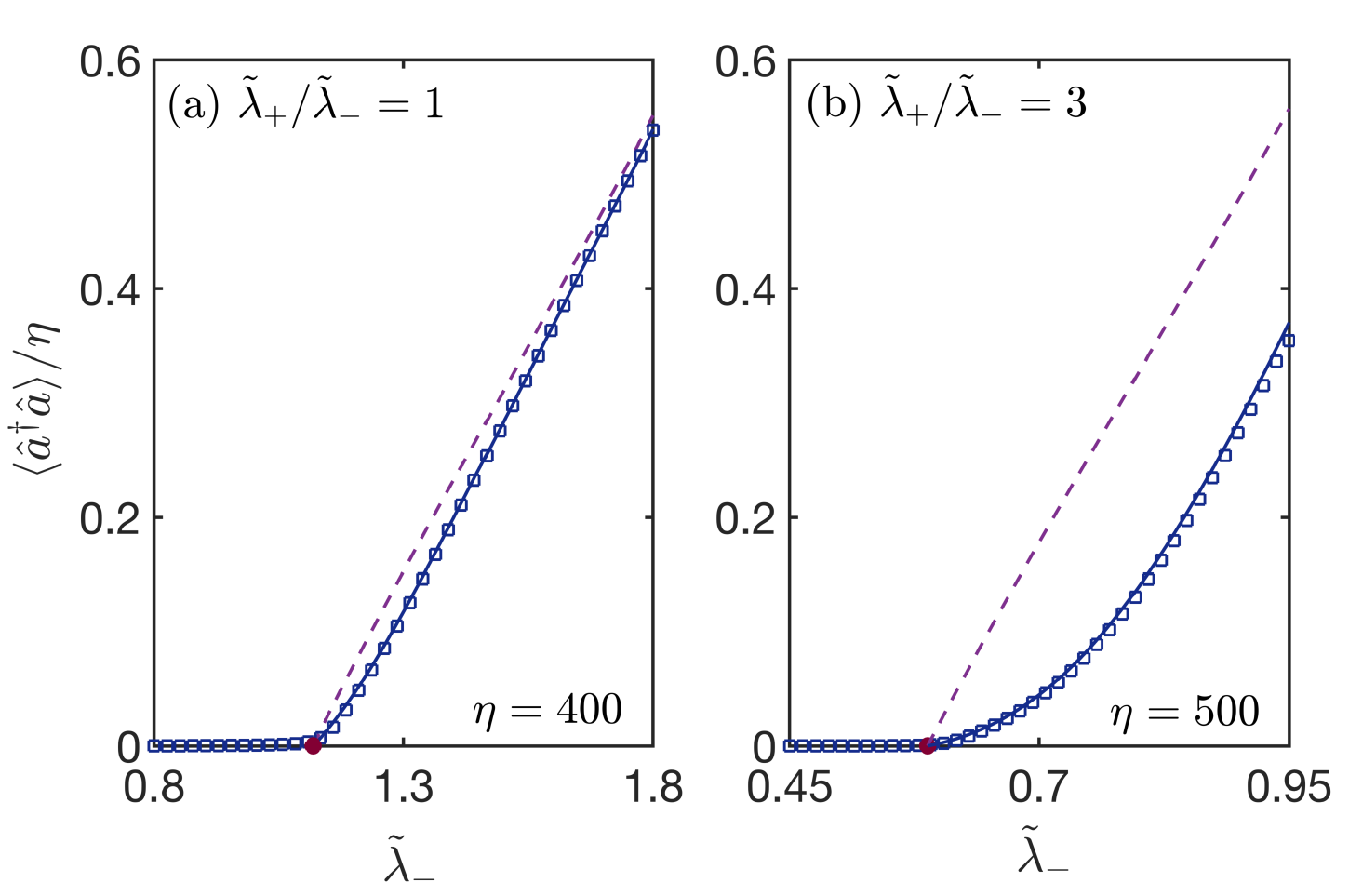}
\caption{\label{fig5-photon} Dependence of $\langle \hat{a}^\dag \hat{a}\rangle/\eta$ on the coupling strength $\tilde\lambda_-$ across the second-order NP-SP transition (the solid critical line of Fig.~\ref{fig2-phase}). Squares are exact numerical data, obtained by solving Eq.~(\ref{eq-me}). The solid curves are the analytical treatment, see Eq.~(\ref{nSR}). The dashed curves are the order parameter $|c|^2$ of the pure ${\rm SP}_\downarrow$ state, see Eq.~(\ref{eq-sp}). In panel (a) we have used $\tilde\lambda_+/\tilde\lambda_-=1$ and $\eta=400$ while in panel (b) $\tilde\lambda_+/\tilde\lambda_-=3$ and $\eta=500$. Other parameters are the same as Fig.~\ref{fig1-phase}.}
\end{figure}

The value of $ P_{{\rm SP}_\downarrow}$ has a direct effect on the order parameter of the second-order phase transition. The average photon number in the superradiant phase is:
\begin{equation}\label{nSR}
\langle \hat{a}^\dag \hat{a} \rangle = P_{{\rm SP}_\downarrow} |c|^2 \eta,
\end{equation}
where $c$, given by Eq.~(\ref{eq-sp}), is for the pure ${\rm SP}_\downarrow$ state. In Fig.~\ref{fig5-photon} we show a comparison of Eq.~(\ref{nSR}) to the numerical results obtained by solving the master equation at large $\eta$, finding excellent agreement. The admixture with ${\rm NP}_\uparrow$ can cause a significant reduction from the photon number of the pure superradiant state (dashed curves). The discrepancy is more obvious close to the phase transition, where $\theta_{\rm SP} \simeq 0$. Instead, deep in the superradiant phase the spin direction moves away from the $z$ axis, which causes a reduction of $\cos\theta_{\rm SP}$ and a growth of $P_{{\rm SP}_\downarrow}$ [see Eq.~(\ref{Pm_SR})]. As a result, we see in Fig.~\ref{fig5-photon}(a) that the exact photon number tends to approach the dashed curve by increasing $\tilde\lambda_-$. Furthermore, close to the critical point we  have $|\tilde\chi_+|\simeq \tilde\lambda_+$, thus we can approximate $P_{{\rm SP}_\downarrow} \simeq (1+\lambda^2_+/\lambda^2_-)^{-1}$. This expression indicates that a larger value of $\lambda_+/\lambda_-$ leads to a more dramatic reduction of $\langle \hat{a}^\dag \hat{a} \rangle$, in agreement with panel (b). The physical interpretation is quite clear: Strong counter-rotating terms can excite simultaneously the qubit and the cavity. This, in conjunction with cavity decay, drives a large population to the NP$_\uparrow$ high-energy state, suppressing superradiance.

\subsection{First-order phase transition}\label{sec:Cphase}

We now examine the coexistence region C of Fig.~\ref{fig1-phase}, where the stationary state is expected to be a statistical mixture of the ${\rm NP}_\uparrow$, ${\rm NP}_\downarrow$, and ${\rm SP}_\downarrow$. Remarkably, instead, we find that in region C the system returns to the same normal state described before, i.e., the populations of ${\rm NP}_{\uparrow/\downarrow}$ are as in Eq.~(\ref{Ppm}) and $P_{{\rm SP}_\downarrow}=0$. Therefore, the phase diagram for the stationary states (i.e., taking $t\to \infty$ before $\eta\to \infty$) becomes as in Fig.~\ref{fig2-phase}, where the mean-field phase boundary between the superradiant (SP) and C regions has now become a first-order transition between SP and NP.

\begin{figure}
\centering
\includegraphics[width=0.48\textwidth]{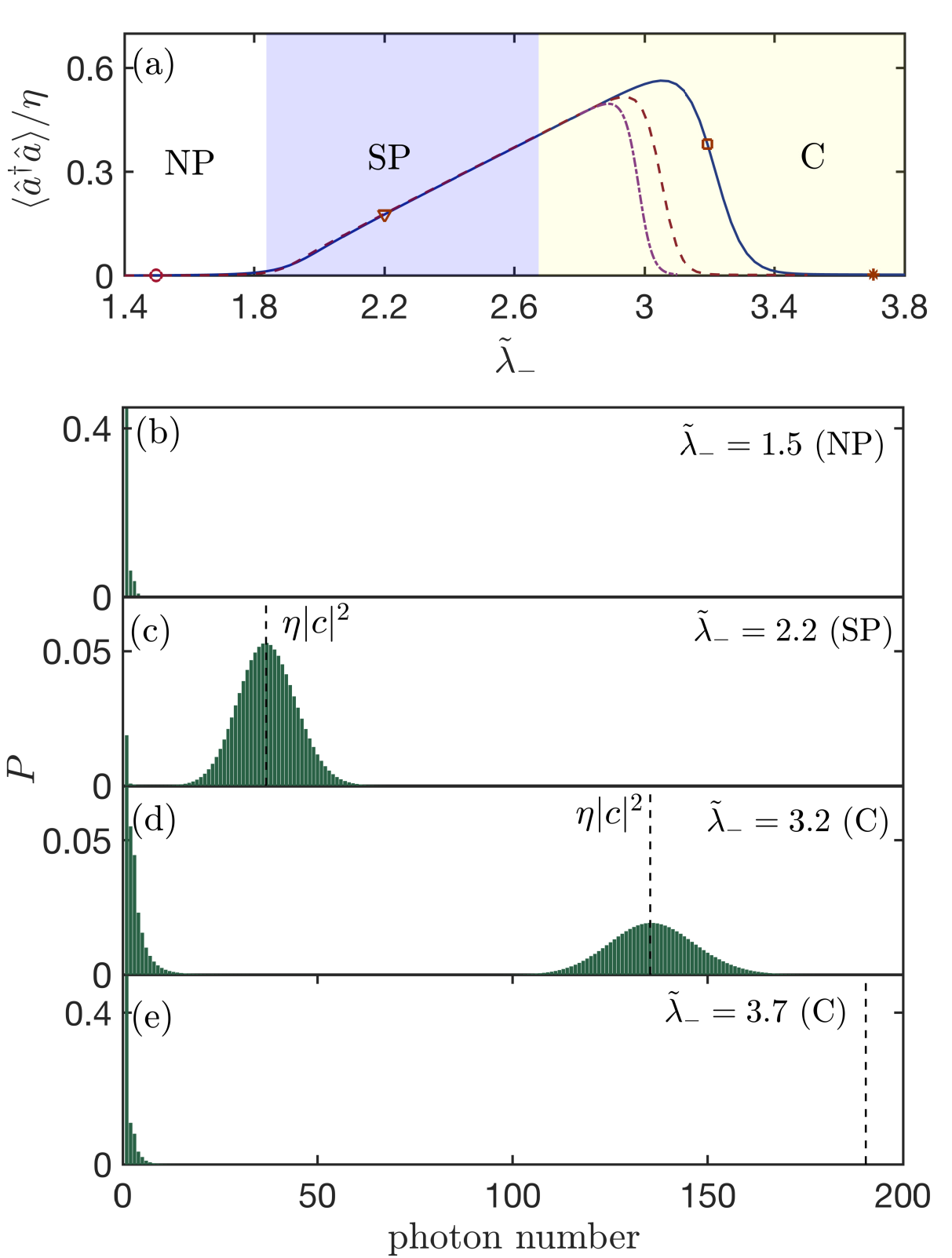}
\caption{\label{fig6} Dependence of the photon number on the coupling strength. In panel (a) we plot $\langle \hat{a}^\dag \hat{a}\rangle/\eta$ as a function of $\tilde\lambda_-$. The three curves are for $\eta = 200$ (solid), 400 (dashed), and 600 (dot-dashed). In panels (b-e) we show the detailed photon distribution with $\eta=200$ and four representative values of $\tilde\lambda_-$, marked by symbols in panel (a). Panel (b) and (c) are respectively in the normal and superradiant phases. Panels (d) and (e) are both in the `coexistence' phase C of Fig.~\ref{fig1-phase}. However, a significant mixture of normal and superradiant states only occurs for panel (d). At large $\tilde\lambda_-$ the system returns to the normal state, see panel (e). The vertical dashed lines in panels (c-e) indicate the mean-field value of $\langle \hat{a}^\dag \hat{a}\rangle$ for the superradiant state, given by Eq.~(\ref{eq-sp}). In all panels $\tilde\lambda_+ = 0.3 \tilde\lambda_-$ and $\kappa=0.5\omega_0$.
}
\end{figure}

In Fig.~\ref{fig6}, we show evidence of this behavior from numerical simulations based on the master equation. We consider relatively large values of $\eta$ and compute the average photon number as a function of $\tilde\lambda_-$ (at a fixed ratio $\tilde\lambda_+/\tilde\lambda_- =0.3$), finding that the monotonic increase of $\langle \hat a^\dag \hat a\rangle$ breaks down after entering the C region. As seen in Fig.~\ref{fig6}(a), there is a sudden drop at sufficiently large $\tilde\lambda_-$, bringing the system close to $\langle \hat a^\dag \hat a\rangle/\eta \approx 0$. By increasing $\eta$, the drop occurs at progressively smaller values of $\tilde\lambda_-$ and becomes sharper, suggesting the occurrence of a first-order transition when $\eta \to \infty$.

The various regimes can also be studied through the photon number distribution in Fock space, shown in panels (b-e) of Fig.~\ref{fig6} at four representative values of $\tilde\lambda_-$. The behavior in panels (b) and (c) is as expected: in the former (latter) case a prominent peak around $\langle \hat a^\dag \hat a\rangle=0$ ($\langle \hat a^\dag \hat a\rangle=\eta|c|^2$) is found, corresponding to the NP (SP$_\downarrow$) mean-field solution. Note that in panel (c), which should generically show a mixture of NP$_\uparrow$ and SP$_\downarrow$ states, there is no visible peak around zero photons. The small population of ${\rm NP}_\uparrow$ is due to large value of $\tilde\lambda_-$, which is far from the second-order critical point, and the small ratio $\tilde\lambda_+/\tilde\lambda_- = 0.3$ (see the discussions at the end of Sec.~\ref{sec:second_order_transition}). In panel (d) of Fig.~\ref{fig6} we show a state which is clearly a nontrivial mixture of superradiant and normal states. The photon distribution has two prominent peaks, but this behavior is only found in the narrow transition region where $\langle \hat a^\dag \hat a\rangle$ drops. By increasing $\tilde\lambda_-$, the superradiant peak progressively disappears and the system eventually returns to the normal state, see panel (e).

\begin{figure}
\centering
\includegraphics[width=0.45\textwidth]{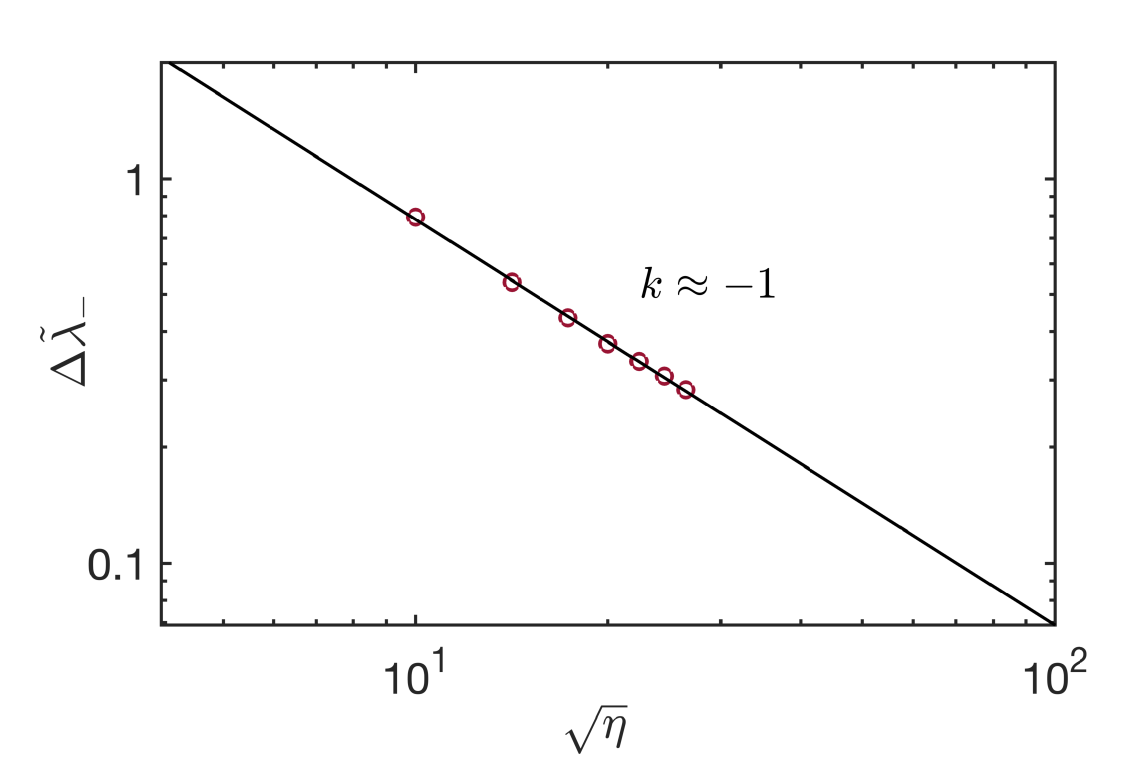} 
\caption{ \label{fig7} Dependence of $\Delta \tilde\lambda_ -= \tilde\lambda_-^* - \tilde\lambda_c^+$ on $\eta$.  Here $\tilde\lambda_-^* $ is the position at which there is a rapid drop in photon number, see Fig.~\ref{fig6}(a). It is found from the inflection points, i.e., by imposing $d^2\langle \hat a^\dag \hat a\rangle/d\tilde\lambda_-^2=0$. The solid line shows a power-law dependence of type $1/\sqrt{\eta}$. The parameters are as in Fig.~\ref{fig6}.
}
\end{figure}

Unfortunately, the limit $\eta \to \infty$ is difficult to approach through exact numerical simulations. Due to the large photon number in the SP phase (where  $\langle \hat a^\dag \hat a\rangle \propto \eta$), computing accurately the drop of $\langle \hat a^\dag \hat a\rangle $ requires a large Hilbert space. This point can be also appreciated from panels (b-e) of Fig.~\ref{fig6}. At the largest value of $\eta$ shown in Fig.~\ref{fig6}(a), convergence to the $\eta \to \infty$ limit has not yet been achieved. To infer the position of the critical point we plot in Fig.~\ref{fig7} the dependence on $\eta$ of $\Delta \tilde\lambda_ -= \tilde\lambda_-^* - \tilde\lambda_c^+$, where $\tilde\lambda_-^*$ marks the inflection point of $\langle \hat a^\dag \hat a\rangle$ and $\tilde\lambda_c^+$ is the SP/C boundary, given in Eq.~(\ref{lambdac_pm}). We find a dependence of the type $\Delta \tilde\lambda_- \sim \eta^{-1/2}$, indicating that the critical point is indeed at $\tilde\lambda_c^+$.

\begin{figure}
\centering
\includegraphics[width=0.45\textwidth]{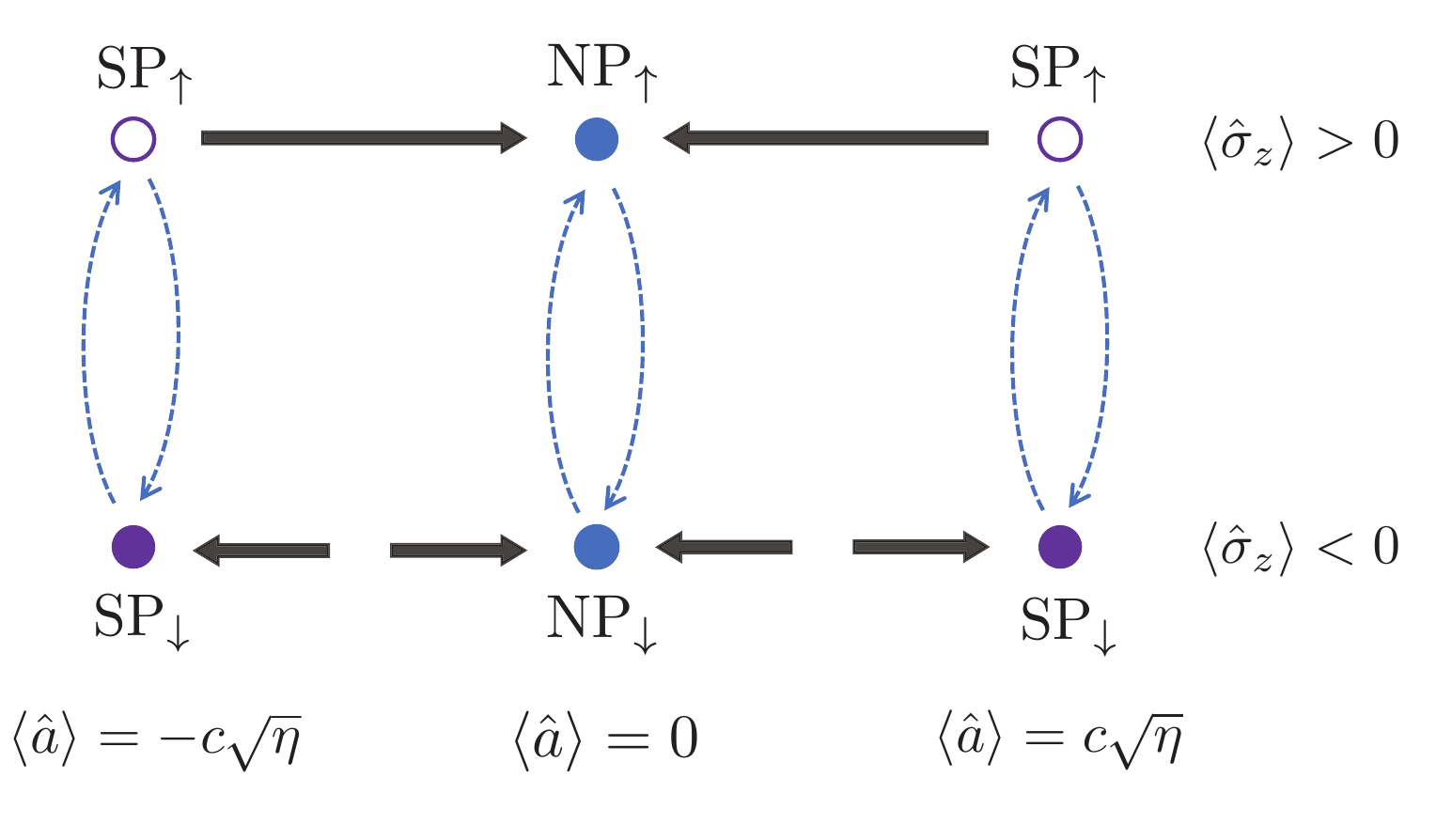}
\caption{\label{schematics_rates_2} Relevant transitions taking place in the C region. The notation is the same of Fig.~\ref{schematics_rates_1} but now also ${\rm NP}_\downarrow$ is a stable fixed point, with its own basin of attraction. }
\end{figure}


The above numerical analysis is consistent with a physical argument based on effective master equations. Similarly to Fig.~\ref{schematics_rates_1}, we represent in Fig.~\ref{schematics_rates_2} the transitions between mean-field solutions in the C region. As discussed before, the slow process ${\rm SP}_\downarrow \to {\rm SP}_\uparrow$ allows the transition pathway ${\rm SP}_\downarrow \to {\rm SP}_\uparrow \to {\rm NP}_\uparrow $ and, like in the superradiant phase, the metastable state ${\rm NP}_\uparrow$ undergoes the slow transition ${\rm NP}_\uparrow  \to {\rm NP}_\downarrow $.  However, differently from  Fig.~\ref{schematics_rates_1}, ${\rm NP}_\downarrow $ is now a stable fixed point, thus the process $ {\rm NP}_\downarrow \to {\rm SP}_\downarrow $ is not available. Once the system has reached one of the two normal states, it will perform ${\rm NP}_\uparrow \leftrightarrow {\rm NP}_\downarrow $ transitions with the same rates of Eqs.~(\ref{Gamma_NP1}) and (\ref{Gamma_NP2}) and will never return to the superradiant state. Therefore, we find the same statistical mixture of the normal phase, with no discontinuity at the dashed boundaries of Fig.~\ref{fig1-phase}. The main predictions of Fig.~\ref{schematics_rates_2}, i.e., the occurrence of a first-order phase transition at $\tilde\lambda_- = \tilde\lambda_c^+$ (dotted lines in Fig.~\ref{fig1-phase}) and the absence of a superradiant contribution in the C region, are in agreement with our numerical findings.

\section{Scaling analysis}
\label{finite frequency scaling}

In this section we return to the second-order phase transition between the NP and SP and perform a finite-$\eta$ scaling analysis based on an effective equilibrium theory. This approach was originally proposed for the dissipative Dicke model~\cite{dalla2013keldysh}, where it shows good agreement with the more rigorous Keldysh path-integral approach~\cite{dalla2013keldysh,nagy2015nonequilibrium,sieberer2016keldysh}. Here, since the phase transition is due to the $\downarrow$ subspace, we initially assume $\hat\sigma_z = -1$. We need, however, to go beyond the quadratic approximation of the effective Hamiltonian adopted so far and include the leading-order nonlinear terms. With the quadratures $\hat x=(\hat a^\dag+\hat a)/\sqrt{2}$ and $\hat{p}=i(\hat a^\dag-\hat a)/\sqrt{2}$, we find (see Appendix~\ref{app-e-me}):
\begin{align}\label{Heff_xp}
\hat{H}_{eff} \simeq \omega_c \left[\frac{\alpha_y}{2}  \hat{p}^2 + \frac{\alpha_x}{2} \hat{x}^2 +\frac{(\tilde\lambda_x^2\hat{x}^2 +\tilde\lambda_y^2\hat{p}^2 )^2}{16\eta}  \right],
\end{align}
where the rescaled coupling constants $\tilde\lambda_{x,y} = (\tilde\lambda_+ \pm \tilde\lambda_-)/2$ correspond to Eq.~(\ref{lambda_xy}). Furthermore:
\begin{equation}
\alpha_{x,y} =1 - \tilde\lambda_{x,y}^2.
\end{equation} 
The quadratic part of Eq.~(\ref{Heff_xp}) is in agreement with $\hat{H}_{np}$ of Eq.~(\ref{Hnp}) and the additional terms are suppressed by a factor of $1/\eta$, thus they can be neglected in the normal phase. These quartic terms, however, become crucial in the critical region since $\langle \hat{x}^2\rangle \sim \eta$ in the superradiant phase. At variance with the closed system \cite{liu2017universal}, both $\langle \hat{x}^2\rangle$ and $\langle \hat{p}^2\rangle$ scale linearly with $\eta$ in the superradiant phase. Therefore, we cannot neglect $\tilde\lambda_y^2 p^2$ in the quartic term.

To compute the critical properties, we consider the following equation of motion:
\begin{equation}\label{damped_oscillator_quantum}
\frac{d^2\hat{x}}{dt^2}= -\omega_c^2 \left(\alpha^2 \hat x +\frac{\beta^3}{\eta} \hat{x}^3\right)-2\kappa \frac{d\hat{x}}{dt}+\hat{\mathcal{G}},
\end{equation}
which can be derived from the Heisenberg-Langevin equations associated to Eq.~(\ref{Heff_xp}). Details are presented in Appendix~\ref{app:Langevin_eq}, where we obtain:
\begin{align}
\alpha^2 = \alpha_x \alpha_y +\frac{\kappa^2}{\omega_c^2},  \quad
\beta^3= \alpha_y\tilde\lambda^4_x 
- \frac{1}{\alpha_y^3}\left(\frac{\kappa\tilde\lambda_y}{\omega_c}\right)^4 , \label{alpha_beta}
\end{align}
and the expression of the noise operator $\hat{\mathcal{G}}(t)$, given in Eq.~(\ref{G_noise}). While Eq.~(\ref{damped_oscillator_quantum}) is an operator equation, we follow the idea that the critical behavior is determined by classical fluctuations \cite{dalla2013keldysh}. Then, an accurate description of the critical behavior can be obtained from the classical limit of Eq.~(\ref{damped_oscillator_quantum}), describing a Brownian particle. The potential is:
\begin{equation}\label{V}
V(x) = \omega_c\left(\frac12 \alpha^2 x^2 +\frac{1}{4\eta}\beta^3 x^4 \right),
\end{equation}
which is in agreement with the mean-field treatment, since the second-order phase boundary is recovered by setting $\alpha^2 = 0$ and the equilibrium position $x = -\alpha^2/\beta^3$ is consistent with the mean-field value of $c$ in Eq.~(\ref{eq-sp}). In Appendix~\ref{app:Langevin_eq} we also derive the effective temperature associated with the noise operator $\hat{\mathcal{G}}$:
\begin{equation}\label{Teff_main}
T = \frac{\kappa^2 +\alpha_y^2\omega_c^2}{4\omega_c}. 
\end{equation}
 
To compute the critical properties, we should finally recall that the above classical limit only refers to the $\hat\sigma_z = -1$ subspace. The fluctuations in the $\hat\sigma_z = +1$ subspace do not diverge with $\eta$, so the contribution of these states is negligible. This implies that a prefactor $P_-$ (i.e., the population of $\hat\sigma_z = -1$) should be included when computing the expectation values of $\langle \hat x^2 \rangle$ by a classical average:
\begin{align}\label{statistical_average}
\langle \hat x^2 \rangle =  & P_- \frac{\int  x^2e^{-V(x)/T}dx}{\int e^{-V(x)/T}dx} \nonumber \\
& = 2P_- \sqrt{\frac{\eta T}{\beta^3 \omega_c}}Q\left(-\frac{\alpha^2}{2}  \sqrt{\frac{\eta \omega_c}{\beta^3 T}}\right).\end{align}
The scaling form in the second line of Eq.~(\ref{statistical_average}) is obtained after a simple change of variable in the integrals, giving:
\begin{equation}\label{Q}
Q(x) = \frac{\int z^2 e^{-(x-z^2)^2}dz}{\int e^{-(x-z^2)^2}dz}.
\end{equation}
It is also useful to linearize $\alpha^2$ close to the critical point. By considering the line $\tilde\lambda_+ = r \tilde\lambda_-$, we expand $\alpha^2$ in terms of $\delta\tilde\lambda_- = \tilde\lambda_- - \tilde\lambda_c^-$, where the critical point $\lambda_c^-$ is given by Eq.~(\ref{lambdac_pm}) and:
\begin{align}\label{alpha2_linear}
\alpha^2  \simeq  - \tilde\lambda_c^-\left[1+r^2 - \frac{(1-r^2)^2}{4}(\tilde\lambda_c^{-})^2 \right] \delta\tilde\lambda_- 
\equiv -\Lambda \delta\tilde\lambda_-.
\end{align}
From Eq.~(\ref{statistical_average}), we can also easily obtain the expression for the average photon number $\langle \hat{a}^\dag \hat{a} \rangle $, using the approximate relation $\hat{p} \simeq \kappa \hat{x}/\alpha_y \omega_c$ (see Appendix~\ref{app:Langevin_eq}).

\subsection{Critical fluctuations}

\begin{figure}
\centering
\includegraphics[width=0.48\textwidth]{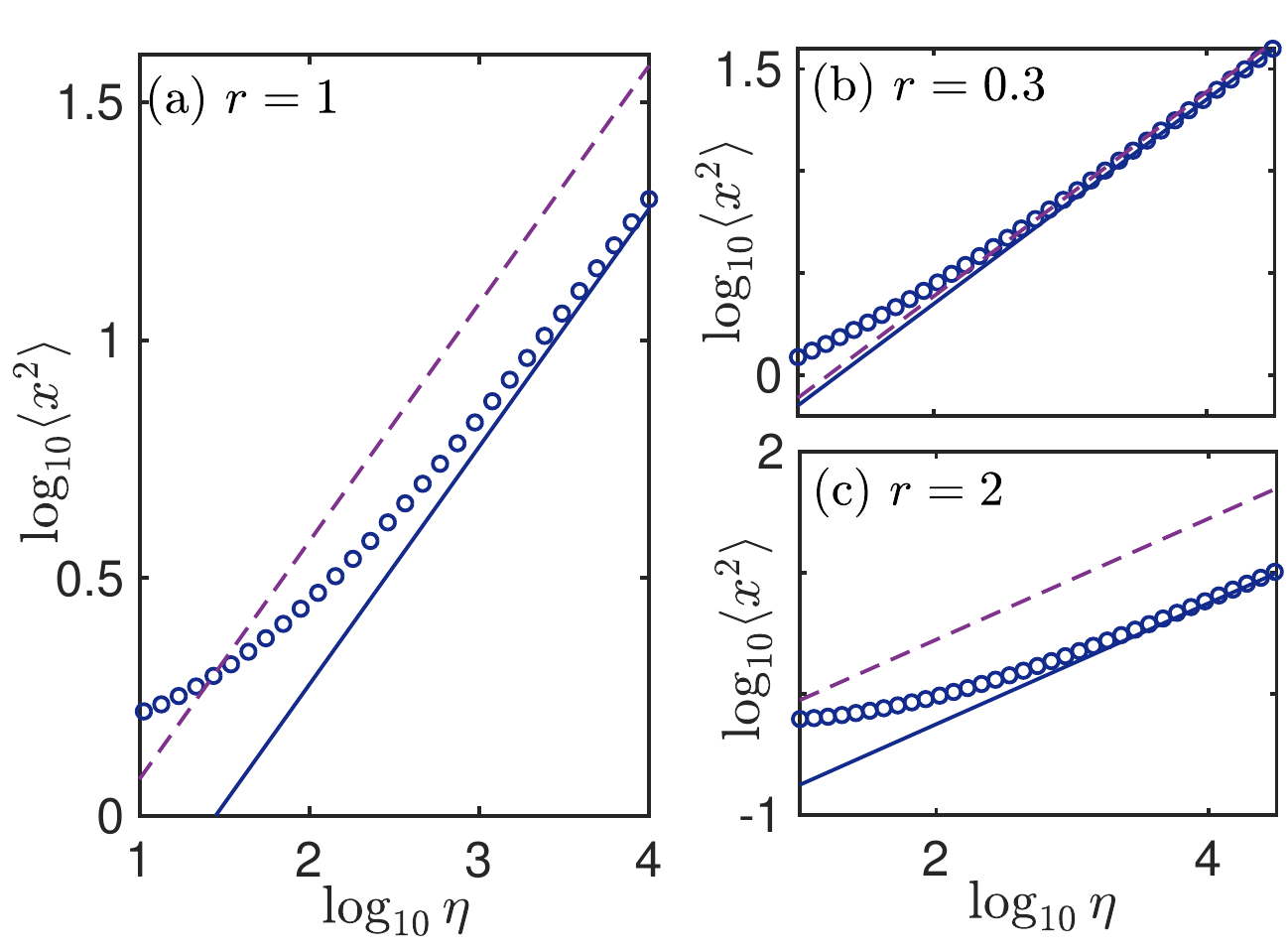}
\caption{\label{fig9-x2} Dependence of $\langle \hat{x}^2 \rangle$ on $\eta$ at different points on the second-order phase boundary. The critical points are parameterized by $r$, see Eq.~(\ref{lambdac_xy}), and we use $r=1,0.3,2$ in panels (a,b,c), respectively. The blue circles are obtained from the master equation while the solid curves are the analytical prediction of Eq.~(\ref{critical_x2}). The dashed curves omit the $P_-$ prefactor. As explained in the main text, these curves may be referred to the thermodynamic limit of the Dicke model (where $\eta = N \Omega/\omega_c$ \cite{liu2017universal}). We used $\kappa = 0.5 \omega_c$.}
\end{figure}

We now evaluate Eq.~(\ref{statistical_average}) on the second-order phase boundary, which can be parametrized with $r=\tilde\lambda_+ / \tilde\lambda_-$. The fluctuations at the critical point are:
\begin{equation}\label{critical_x2}
\langle \hat{x}^2 \rangle = \left(\frac{2Q(0)}{1+r^2} \sqrt{\frac{T}{\beta^3 \omega_c} }\right) \sqrt{\eta},
\end{equation} 
where $Q(0)=\Gamma(3/4)/\Gamma(1/4) \simeq 0.338$ and the values of $\beta^3$ and $T$ can be computed using:
\begin{equation}\label{lambdac_xy}
\tilde\lambda_x = \frac{r+1}{2}\tilde\lambda_c^-, \qquad 
\tilde\lambda_y = \frac{r-1}{2}\tilde\lambda_c^- ,
\end{equation}
where $\tilde\lambda_c^-$ is given in Eq.~(\ref{lambdac_pm}). Since both $T$ and $\beta^3$ are evaluated using Eqs.~(\ref{lambdac_xy}) and (\ref{lambdac_pm}), the prefactor in Eq.~(\ref{critical_x2}) becomes a function of $r$ and $\kappa/\omega_c$. For the isotropic Rabi model ($r=1$) we find the simpler formula:
\begin{equation}\label{critical_1}
\langle \hat{x}^2 \rangle = \frac{Q(0)}{2}
\sqrt{\frac{\eta }{1+\kappa^2/\omega_c^2}}.    \qquad (r=1)
\end{equation}
In Fig.~\ref{fig9-x2}, we show a comparison of the approximate Eq.~(\ref{critical_x2}) to exact numerical results at three different points on the phase boundary, finding excellent agreement at sufficiently large $\eta$.

The above results give a critical scaling:
\begin{align}
\label{eq23}
\langle x^2\rangle\propto \eta^{\zeta},
\end{align}
with the same exponent $\zeta=1/2$ of the isotropic limit~\cite{hwang2018dissipative}.  This conclusion is hardly surprising, as a finite anisotropy is not expected to modify the universality class of the model. Instead, it is more interesting to perform a detailed comparison to the critical scaling of the Dicke model. The comparison was done numerically in Ref.~\cite{hwang2018dissipative} for the isotropic case and is immediate within the above treatment. In fact, the effective equilibrium theory of the anisotropic Dicke model leads to the same potential and temperature of Eqs.~(\ref{V}) and (\ref{Teff_main}), with the $\eta$ parameter generalized as follows~\cite{liu2017universal}:
\begin{equation}
\eta \to \frac{N \Omega}{\omega_c}.
\end{equation}
Besides this replacement, an important difference is due to spin fluctuations, which introduce $P_-$ in Eq.~(\ref{statistical_average}). Within the effective equilibrium theory, this is the only non-universal factor relating the two models. In particular, our Eq.~(\ref{critical_1}) differs by $P_- =1/2$ from the corresponding result for the isotropic Dicke model~\cite{dalla2013keldysh}. The non-universal factor found in Ref.~\cite{hwang2018dissipative} is 0.507, close to the exact value predicted here. In the presence of a finite anisotropy ($r\neq 1$), the non-universal factor is simply $P_- =1/(1+r^2)$ and determines the difference between dashed and solid curves of Fig.~\ref{fig9-x2}. Therefore, as shown in panel (b), the difference in the scaling behavior of the Dicke and Rabi models should be strongly reduced at small $r$. In conclusion, the above arguments give a physical interpretation of the non-universal factor, which has been identified with the $\downarrow$ spin population. 

\subsection{Scaling functions}

\begin{figure}
\centering
\includegraphics[width=0.48\textwidth]{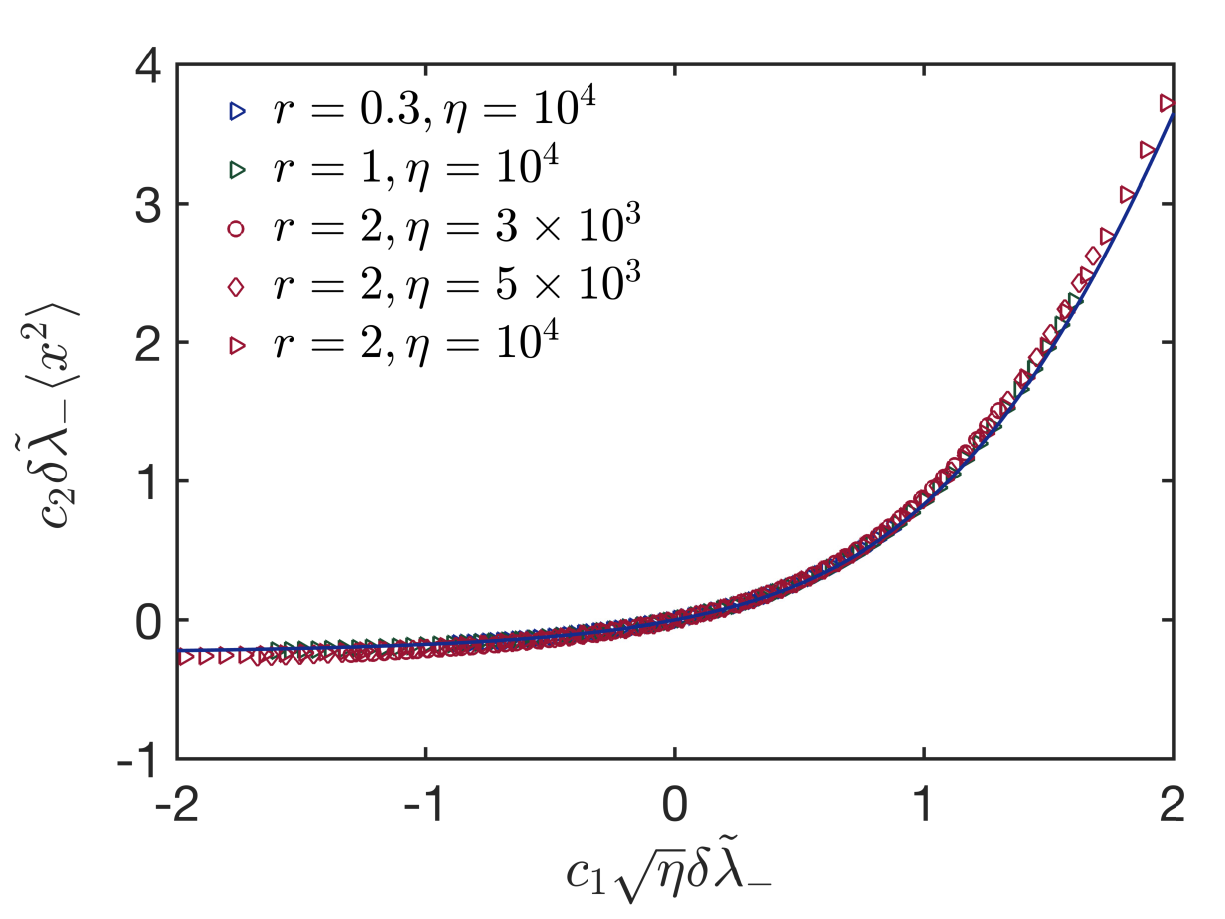}
\caption{\label{fig10_u} Finite-$\eta$ scaling function obtained from the evaluation of $\langle \hat{x}^2 \rangle$. The numerical data (symbols) are computed at different values of $\eta$ and $\tilde\lambda_\pm$ and collapse to a universal function when properly rescaled. The scaling coefficients  $c_{1,2}$ are given in Eq.~(\ref{c12}) and the solid curve is $x Q(x)$, see Eq.~(\ref{Q}). As in previous figures, $\kappa = 0.5 \omega_c$.}
\end{figure}

The effective equilibrium theory also allows us to describe accurately the scaling functions which, as clear from Eq.~(\ref{statistical_average}), can be related to $Q(x)$. Specifically, to make contact with previous literature \cite{hwang2018dissipative}, we first compute from Eq.~(\ref{statistical_average}) the exponent $\nu$, giving the divergence of the fluctuations when approaching the critical point in the limit $\eta \to \infty$:
\begin{equation}
\langle \hat{x}^2 \rangle \simeq P_-  \frac{T}{\omega_c \alpha^2} \propto \frac{1}{|\delta\tilde\lambda_-|^\nu},
\end{equation}
where we assumed $\tilde\lambda_-<\tilde\lambda_c^-$ and used $Q(x)\simeq -\frac{1}{4x}$ when $x \to -\infty$. Expanding $\alpha^2$ as in Eq.~(\ref{alpha2_linear}) we immediately find $\nu =1$. Furthermore, we derive the following scaling relation
\begin{equation}
(\delta\tilde\lambda_-)^\nu \langle \hat{x}^2 \rangle = F \left(\eta (\delta\tilde\lambda_-)^{\nu/\zeta}\right),
\end{equation} 
where $F(x) = (c_1/c_2) \sqrt{x}  Q(c_1 \sqrt{x})$ can be obtained from simple manipulations of Eq.~(\ref{statistical_average}). Here:
\begin{align}\label{c12}
c_1 = \frac{\Lambda}{2} \sqrt{\frac{\omega_c}{\beta^3T}} , \qquad
c_2 = \frac{\Lambda \omega_c}{4P_-T},
\end{align}
where $\Lambda$ is defined by Eq.~(\ref{alpha2_linear}). The above expressions of $c_{1,2}$ allow us to collapse the scaling functions obtained at different values of $r$ along the phase boundary. In Fig.~\ref{fig10_u} we show that the collapse of numerical data is excellent, giving a scaling function which agrees with the analytical prediction of Eq.~(\ref{statistical_average}). Other scaling functions, e.g., describing $\langle \hat{a}^\dag \hat{a}\rangle$, can be obtained in a similar manner.

\section{Discussion}\label{concl}

We have presented a detailed characterization of various dissipative phase transitions of the anisotropic Rabi model with cavity dissipation. A main point brought out by our analysis is the crucial role played by spin fluctuations in determining the properties of the system (e.g., the order parameter), and in reaching a proper understanding of the critical behavior. This finding is unexpected in the context of superradiant phase transitions: In the Dicke model, the transition is realized for $N \to \infty$, i.e., when the system approaches its classical limit. Also the ground-state phase transition of the Rabi model is well understood in terms of small quantum corrections around a `classical oscillator limit'~\cite{BakemeierPRA2012}. Instead, here large quantum fluctuations persist in the stationary state. They lead to a general suppression of superradiance, due to effective incoherent pumping of the qubit, and completely determine the non-universal scaling factors of the second-order transition. Furthermore, superradiance can disappear in certain regions of parameters where, in principle, a mean-field superradiant state would be stable.

We expect the type of processes described here to be generic for dissipative transitions involving few-body systems. Qualitatively, quantum fluctations persists due to the small Hilbert space of the qubit, which makes the high-energy states accessible via a single spin-flip. Therefore, a similar effective spin pumping process would likely play an important role in phase transitions of the Dicke model with fixed $N \geq 2$ (still considering the $\Omega/\omega_c \to \infty$ limit). A wider variety of related few-body transitions, including few-site lattices and central-spin systems, has been discussed in the recent literature~\cite{carmichael2015breakdown,PhysRevX.7.011012,PhysRevLett.117.123602,PhysRevA.101.063843,PhysRevA.107.013714,PhysRevLett.127.063602,PhysRevLett.129.183602,PhysRevLett.130.043602,PhysRevA.107.033702}, and several aspects of the present analysis might be applicable to those systems as well. 

It is also interesting to make contact with other recently discussed non-equilibrium effects. The physics of this model is strictly related to concept of effective spin temperature, which is also important for the dissipative Dicke model (see, e.g., Ref.~\cite{dalla2013keldysh}). In the Rabi model, however, the non-equilibrium state is more extreme since, as mentioned, the effective spin temperature is infinite or negative in half of the parameter space. Therefore, a closer analogy can be perhaps found to non-equilibrium phenomena such as the `breakdown of photon blockade'~\cite{carmichael2015breakdown,PhysRevX.7.011012} of a single spin coupled to a cavity, or the `breakdown phase transition' of the Dicke model~\cite{BreakdownPRL2020}, where the system is led to a highly excited state. Especially, in the `breakdown phase transition', infinite heating of the cavity is induced by a combination of  a strong coherent interaction and qubit decay. The former can simultaneously excite the cavity and the qubits, while dissipation tends to continuously reset the spin state~\cite{BreakdownPRL2020}.

The basic mechanism at play here is the mirror-image of the one described above. Spin pumping is induced by a combined effect of the counter-rotating terms (which can simultaneously excite the qubit and the cavity) and cavity relaxation. However, there are significant differences from the `breakdown phase transition'. Here, for example, a strong collective interaction is not necessary. The effective temperature only depends on the ratio of co- and counter-rotating couplings, thus an infinite or negative spin temperature is realized already at weak coupling.  In general, we expect that dissipative phase transitions of few-body systems may lead more easily to this type of states under extreme non-equilibrium conditions.

Another natural implication of our treatment is that qubit relaxation should be favorable to the formation of the superradiant state. In fact, since the spin-flip rates due to the cavity are suppressed by $\eta$, natural decoherence mechanisms of the qubit eventually dominate when $\eta\to \infty$. This is actually an interesting point, as in more realistic systems the perfect correspondence between the Dicke and Rabi model might be restored (i.e., the non-universal factor $P_-$ could be absent, like in the zero-temperature QPT~\cite{liu2017universal}). On the other hand, the cavity-induced spin-flip rates only decay as $\sim \kappa/\eta$. If the qubit relaxation rate $\gamma$ is much smaller than $\kappa$, a sufficiency large suppression factor might be difficult to achieve (a maximum value $\eta = 25$ has been realized in Ref.~\cite{DuanNatComm2021}). In that case, `finite-size' contributions to the spin-flip transition rates are likely to play an important role in the whole range of accessible parameters. 

Future studies may address more precisely this interplay between qubit and cavity-induced dissipation. It would be also interesting to determine if an enhancement of superrdiance by qubit dissipation could be helpful for recently proposed quantum sensing protocols, based on the critical behavior of the Rabi model~\cite{2020_PRL_Paris_metrology,2021_PRL_sensing,2022_PRXQuantum_sensing,2023_SChina_sensing}
 
\emph{Note added:} During the last stages of preparing this manuscript, we became aware of a related work studying the dissipative phase transitions of the anisotropic quantum Rabi model~\cite{lyu2023multicritical}.

\section{Acknowledgments}

This work has been supported by the ERC under grant agreement No. 101053159 (RAVE), and by the PNRR MUR project PE0000023-NQSTI. Y.D.W. acknowledges support from the NSFC (Grant No. 12275331), the Penghuanwu Innovative Research Center (Grant No. 12047503), and the Innovation Program for Quantum Science and Technology (Grant No. 2021ZD031602). S.C. acknowledges support from the Innovation Program for Quantum Science and Technology (Grant No. 2021ZD0301602), the National Science Association Funds (Grant No. U2230402), and the National Natural Science Foundation of China (Grant Nos. 11974040 and 12150610464). The numerical solutions of the master equation were obtained using QuTiP~\cite{johansson2012qutip,JOHANSSON20131234}.

\appendix

\section{Cumulant expansion}
\label{app-cumulant}
\begin{figure}
\centering
\includegraphics[width=0.45\textwidth]{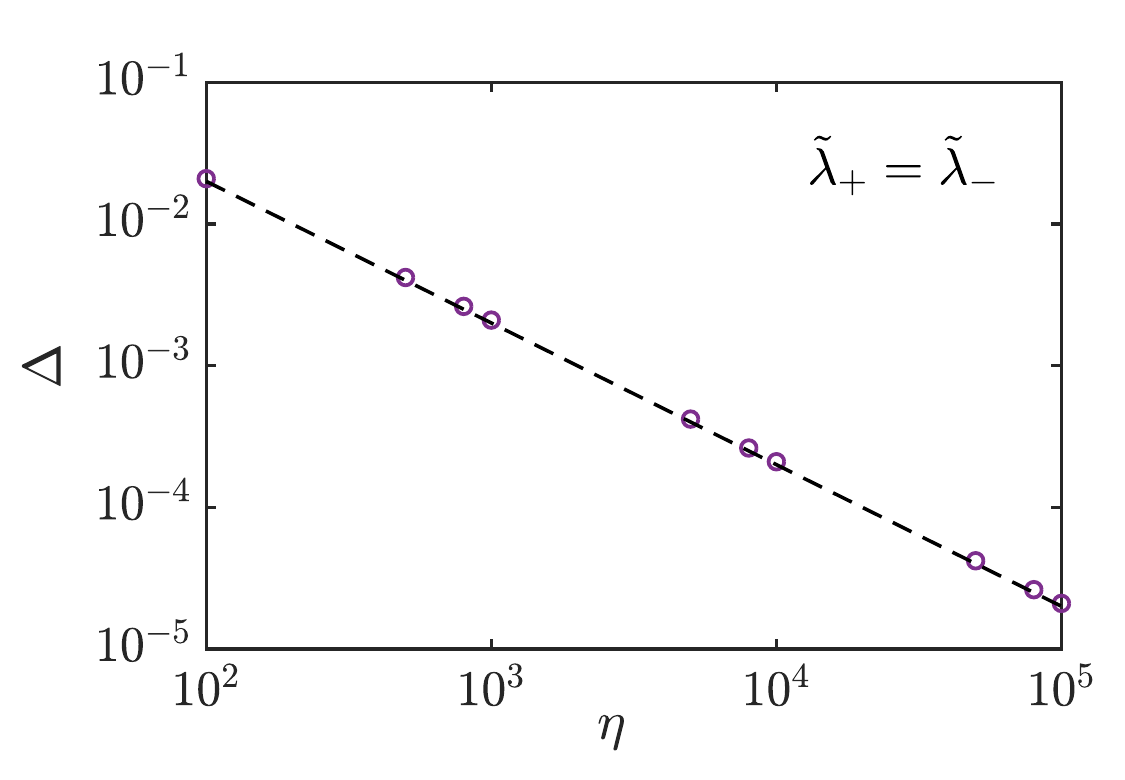}
\caption{\label{fig11_Delta} Comparison of analytical and numerical results for the stationary value of $\langle \hat{\sigma}_z\rangle$. The circles are values of $\langle \hat{\sigma}_z\rangle $ obtained from the exact master equation, see Eq.~(\ref{eq-me}). The dashed line is Eq.~(\ref{Delta}). We used the parameters $\kappa=0.5\omega_c$ and $\tilde\lambda_+=\tilde\lambda_-=0.618$. 
}
\end{figure}

As described in Sec.~\ref{cumulant} of the main text, we have obtained a set of non-linear equations for the second-order correlation functions. Besides Eq.~(\ref{eq-cumul-motion}), we have:
\begin{small}
\begin{align}
\label{eq-B1}
\frac{d\langle \hat{a}^{\dagger}\hat{a}\rangle}{dt}=&-2\kappa\langle \hat{a}^{\dagger}\hat{a}\rangle-i\lambda_-(\langle \hat{a}\hat{\sigma}_{+}\rangle-\langle \hat{a}^{\dagger}\hat{\sigma}_{-}\rangle)\notag \\
&-i\lambda_+(\langle \hat{a}\hat{\sigma}_{-}\rangle 
-\langle \hat{a}^{\dagger}\hat{\sigma}_{+}\rangle),
\\
\frac{d\langle \hat{a}^{2}\rangle}{dt}=&-2\left(\kappa+i\omega_{c}\right)\langle \hat{a}^{2}\rangle+2i\left(\lambda_-\langle \hat{a}\hat{\sigma}_{-}\rangle+\lambda_+\langle \hat{a}\hat{\sigma}_{+}\rangle\right),
\\
\frac{d\langle \hat{a}^{\dagger2}\rangle}{dt}=&-2\left(\kappa-i\omega_{c}\right)\langle \hat{a}^{\dagger2}\rangle-2i(\lambda_-\langle \hat{a}^{\dagger}\hat{\sigma}_{+}\rangle+\lambda_+\langle \hat{a}^{\dagger}\hat{\sigma}_{-}\rangle),
\\
\frac{d\langle\hat{a}\hat{\sigma}_{+}\rangle}{dt}=&
-\left(\kappa+i\omega_c-i\Omega\right)\langle\hat{a}\hat{\sigma}_+\rangle+i\lambda_-\langle\hat{a}^{\dagger}\hat{a} \hat{\sigma}_z\rangle
\notag \\
&+i\lambda_+\langle\hat{a}^{2} \hat{\sigma}_z\rangle
+i\lambda_-\langle\hat{\sigma}_{-}\hat{\sigma}_{+}\rangle, 
\\
\frac{d\langle \hat{a}^{\dagger}\hat{\sigma}_{-}\rangle}{dt}\!=&
\!-\!(\kappa-i\omega_{c}+i\Omega)\langle \hat{a}^{\dagger}\hat{\sigma}_{-}\rangle
-i\lambda_-\langle \hat{a}\hat{a}^{\dagger} \hat{\sigma}_{z}\rangle
\notag \\
&-i\lambda_+\langle \hat{a}^{\dagger2} \hat{\sigma}_{z}\rangle-i\lambda_-\langle\hat{\sigma}_{+}\hat{\sigma}_{-}\rangle,
\\
\frac{d\langle \hat{a}\hat{\sigma}_{-}\rangle}{dt}=&
-(\kappa+i\omega_{c}+i\Omega)\langle \hat{a}\hat{\sigma}_{-}\rangle
-i\lambda_-\langle \hat{a}^{2} \hat{\sigma}_{z}\rangle
\notag \\
&-i\lambda_+\langle (\hat{a}^{\dagger}\hat{a}\hat{\sigma}_{z}\rangle 
-\langle\hat{\sigma}_{-}\hat{\sigma}_{+}\rangle),
\\
\frac{d\langle \hat{a}^{\dagger}\hat{\sigma}_{+}\rangle}{dt}=&
-\left(\kappa-i\omega_{c}-i\Omega\right)\langle \hat{a}^{\dagger}\hat{\sigma}_{+}\rangle
+i\lambda_-\langle \hat{a}^{\dagger2} \hat{\sigma}_{z}\rangle
 \notag \\
&+i\lambda_+(\langle \hat{a}\hat{a}^{\dagger} \hat{\sigma}_{z}\rangle
- \langle \hat{\sigma}_{+}\hat{\sigma}_{-} \rangle),
\\
\frac{d\langle \hat{a}\hat{\sigma}_{z}\rangle}{dt}=&
\!-\!\left(\kappa+i\omega_{c}\right)\langle \hat{a}\hat{\sigma}_{z}\rangle
+2i\left[\lambda_-(\langle\hat{a}^{2}\hat{\sigma}_{+}\rangle
-\langle\hat{a}^{\dagger}\hat{a}\hat{\sigma}_{-}\rangle) \right.
\notag \\
& - \left. \lambda_+(\langle\hat{a}^{2}\hat{\sigma}_{-}\rangle
-\langle\hat{a}^{\dagger}\hat{a}\hat{\sigma}_{+}\rangle)\right]
-i\lambda_-\langle\hat{\sigma}_{-}\rangle
+i\lambda_+\langle\hat{\sigma}_{+}\rangle,
\\
\frac{d\langle \hat{a}^{\dagger}\hat{\sigma}_{z}\rangle}{dt}=&
\!-\!\left(\kappa+i\omega_{c}\right)\langle \hat{a}^{\dagger}\hat{\sigma}_{z}\rangle  
-2i\left[
\lambda_-(\langle\hat{a}^{\dagger2}\hat{\sigma}_{-}\rangle-\langle\hat{a}\hat{a}^{\dagger}\hat{\sigma}_{+}\rangle) \right. \notag \\
&
\left.-\lambda_+(\langle\hat{a}^{\dagger2}\hat{\sigma}_{+}\rangle
-\langle\hat{a}\hat{a}^{\dagger}\hat{\sigma}_{-}\rangle) \right]
+i\lambda_-\langle\hat{\sigma}_{+}\rangle-i\lambda_+\langle\hat{\sigma}_{-}\rangle.
\end{align}
\end{small}
In the above equations the third-order correlators appear explicitly. As explained in the main text, we treat them by neglecting the third-order moments and obtain a closed set of differential equations. Considerable simplifications occur in the normal phase, where we can set $\langle \hat{a}\rangle = \langle\hat{a}^{\dagger}\rangle=\langle \hat{\sigma_{\pm}}\rangle=\langle \hat{a}\hat{\sigma}_z\rangle=\langle \hat{a}^{\dagger}\hat{\sigma}_z\rangle = 0$. Using this condition, and discarding higher-order terms of the large-$\eta$ expansion, we obtain Eq.~(\ref{eq-cumul}) of the main text, where $\Delta$ is:
\begin{align}\label{Delta}
\Delta=  \frac{ 
4\lambda_-^2\lambda_+^2\left[4\omega_c^2\lambda_{\rm diff}^2
-2\omega_c\Omega\left(\kappa^2+\omega_c^2\right) \right]}
{\lambda_{\rm diff}^8
-(\lambda_-^2+\lambda_+^2)^2
[2\omega_c\Omega \lambda_{\rm diff}^2-\Omega^2(\kappa^2+\omega_c^2)]}.
\end{align}
Above, we defined $\lambda_{\rm diff}^2 = \lambda_-^2- \lambda_+^2$. One can see that this expression is of order $1/\eta$ by rescaling the couplings as $\lambda_\pm = \tilde\lambda_\pm \lambda_c$, where $\tilde\lambda_\pm $ are of order unity and $\lambda_c =\sqrt{\Omega\omega_c}/2$. In Fig.~\ref{fig11_Delta} we perform a comparison of Eq.~(\ref{Delta}) with the subleading correction to $\langle \hat{\sigma}_z\rangle$, obtained from a direct solution of the master equation. As seen, there is good agreement between numerical and analytical results.

\section{Effective master equations}
\label{app-e-me}

We give here some additional details on the effective master equation approach. First we consider how the Hamiltonian is transformed.  In the normal phase, the appropriate unitary transformation $\hat{U}_{np}$ is given by Eq.~(\ref{Unp}) which, to second order in the coupling parameters, leads to Eq.~(\ref{Hnp}) of the main text. That effective Hamiltonian, besides being diagonal with respect to $\hat{\sigma}_z$, is quadratic in the bosonic operators. To include nonlinear effects due to the spin-cavity interaction, higher-order terms of $\hat{H}_{np}$ should be computed, which are rather cumbersome to obtain directly using $\hat{U}_{np}$. It is more convenient to rely on the method developed in Ref.~\cite{liu2017universal}, from which the following effective Hamiltonian is obtained, in terms of the quadratures $\hat x=(\hat a^\dag+\hat a)/\sqrt{2}$ and $\hat{p}=i(\hat a^\dag-\hat a)/\sqrt{2}$:
\begin{align}\label{Heff_sqrt}
\hat{H}_{eff}=\omega_c \hat{a}^\dag \hat{a} 
+ \frac{\Omega}{2}\hat\sigma_z 
\sqrt{ 1+\frac{2}{\eta}\left[\tilde\lambda_y^2 \hat{p}^2  
 + \tilde\lambda_x^2 \hat{x}^2 - \tilde\lambda_x \tilde\lambda_y \hat\sigma_z\right] }.
\end{align}
Here, to simplify the notation, we define the couplings  $\tilde\lambda_x = (\tilde\lambda_+ + \tilde\lambda_-)/2$ and $\tilde\lambda_y = (\tilde\lambda_+ - \tilde\lambda_-)/2$. 
Expanding the square root of Eq.~(\ref{Heff_sqrt}) in a Taylor series, and setting $\hat\sigma_z = -1$, we immediately obtain Eq.~(\ref{Heff_xp}) of the main text. In Eq.~(\ref{Heff_xp}) we have omitted a constant, as well as negligible quadratic terms $\propto 1/\eta$.

For the superradiant states, the transformation of the Hamiltonian is more involved. In the following, we assume the mean-field solution $\langle \hat{a} \rangle = \sqrt{\eta} c $, while the other broken-symmetry state (with $c \to -c$) can be treated in a completely analogous manner. As described in the main text, first a displacement transformation $\hat{U}_d$ is applied to the photon field, to yield $\langle \hat{a}^\dag \hat{a} \rangle \simeq 0$. This step leads to the effective Hamiltonian (\ref{Hd}), given in the main text. Next, the following $\hat{U}_{sp}$ is applied:
\begin{align}
\label{eq-C11}
\hat{U}_{sp}=\exp\left[\frac{\cos\theta_{\rm SP}}{2\sqrt{\eta}}\left(\tilde\chi_-\hat{a}\hat{\tau}_+ +\tilde\chi_+\hat{a}^{\dagger}\hat{\tau}_+ - {\rm H. c.}\right)\right],
\end{align}
which is analogous to the $\hat{U}_{np}$ transformation given in Eq.~(\ref{Unp}). Here $\tilde \chi_\pm$ corresponds to $\tilde\lambda_\pm$ and the appearance of $\cos\theta_{\rm SP}$ can be attributed to the modified spin-splitting in Eq.~(\ref{Hd}). A direct evaluation of the transformed Hamiltonian gives:
\begin{align}\label{Hsp_complete}
\hat{U}_{sp}^\dag\hat{H}_{d} & \hat{U}_{sp} \simeq \hat{H}_{sp} -\frac{\omega_c}{2}\cos\theta_{\rm SP}\Big[\tilde\chi_z
\hat{a}^\dag \hat{a}(\tilde\chi_+ \hat{\tau}_+ + \tilde\chi_-^* \hat{\tau}_-) \nonumber \\
&+ \tilde\chi_z
\hat{a}^2(\tilde\chi_+^* \hat{\tau}_- + \tilde\chi_- \hat{\tau}_+)
+\tilde\chi_z \tilde\chi_+ \hat\tau_+
+{\rm H.c.}\Big].
\end{align}
where we neglected terms of order $\omega_c/\sqrt{\eta}$. The main contribution to Eq.~(\ref{Hsp_complete}) is $\hat{H}_{sp}$, given in Eq.~(\ref{Hsp}) of the main text. Here we also include a correction to $\hat{H}_{sp}$ which is not diagonal in $\hat{\tau}_z$ and originates from the spin-dependent drive of $H_d$ [i.e., the last term in the first line of Eq.~(\ref{Hd})]. Such type of terms does not appear in the normal-phase Hamiltonian $\hat{H}_{np}$ which, to the same order of approximation, is diagonal in $\hat\tau_z$. 

Despite this difference between the normal and superradiant states, the off-diagonal terms of Eq.~(\ref{Hsp_complete}) are negligible for our discussions. First we note that they are much smaller than the original off-diagonal terms (which in $\hat{H}_d$ were of order $\sqrt{\eta}\omega_c$). At the same time, the gap of $\hat{H}_{sp}$ between the $\hat{\tau}_z = \pm 1$ states is large (it is approximately $\Omega = \eta \omega_c$). Therefore, if desired, we could explicitly eliminate the off-diagonal corrections by applying a second unitary transformation $\hat U'_{sp} = \exp[\hat S']$, where $\hat S'\sim 1/\eta$. It is then not difficult to check that the dominant part of the Hamiltonian, $\hat{H}_{sp}$, would remain unchanged and all the higher-order corrections would be at most of order $\omega_c/\sqrt{\eta}$. This argument justifies omitting in Eq.~(\ref{Hsp}) the off-diagonal term of Eq.~(\ref{Hsp_complete}).

After having obtained Eqs.~(\ref{Hnp}) and (\ref{Hsp}) as the appropriate effective Hamiltonians, we discuss the effect of dissipation. Before applying $\hat{U}_{np}$ and $\hat{U}_{sp}$, the master equations have a similar form, i.e., Eq.~(\ref{eq-me}) in the original frame and
\begin{align}
\label{meq_rhod}
\dot{\hat{\rho}}_d=
-i\left[ \hat{H}_{d},\hat{\rho}_d\right]
+\kappa\mathcal{D}[\hat{a}]\hat{\rho}_d,
\end{align}
after the displacement transformation, where $\hat{\rho}_d=\hat{U}_{d}^\dag\hat{\rho} \hat{U}_{d}$. We first discuss the master equation for the normal states where, after applying $\hat{U}_{np}$ to Eq.~(\ref{eq-me}) and neglecting corrections smaller than $\kappa/\eta$, we get:
\begin{align}
\label{dissipator_complete}
\dot{\hat{\widetilde{\rho}}}\simeq & -i\left[\hat{H}_{np},\hat{\widetilde{\rho}}\right]+\kappa\mathcal{D}[\hat{a}]\hat{\widetilde{\rho}}
  \notag \\
&-\frac{\kappa}{2\sqrt{\eta}}\left(
\mathcal{D}_C[\hat{a},\hat{O}]\hat{\widetilde{\rho}}
+\mathcal{D}_C[\hat{O}^\dagger,\hat{a}^{\dagger}]\hat{\widetilde{\rho}}\right) \notag \\
&-\frac{\kappa}{4\eta}\tilde{\lambda}_-\tilde{\lambda}_+\left(\mathcal{D}_C[\hat{\sigma}_+,\hat{\sigma}_+]\hat{\widetilde{\rho}}
+\mathcal{D}_C[\hat{\sigma}_-,\hat{\sigma}_-]\hat{\widetilde{\rho}}\right) \notag \\
&+\frac{\kappa}{4\eta}\left(\tilde{\lambda}_-^2 - \tilde{\lambda}_+^2\right)\left(\mathcal{D}_C[\hat{a},\hat{a}]\hat{\widetilde{\rho}}
+\mathcal{D}_C[\hat{a}^{\dagger},\hat{a}^{\dagger}]\hat{\widetilde{\rho}}\right) \notag \\
&+\frac{\kappa}{4\eta}\left(
\tilde{\lambda}_-^2\mathcal{D}[\hat{\sigma}_-]\hat{\widetilde{\rho}}
+\tilde{\lambda}_+^2\mathcal{D}[\hat{\sigma}_+]\hat{\widetilde{\rho}}\right)
.
\end{align}
In Eq.~(\ref{dissipator_complete}), $\mathcal{D}_C[\hat{O}_1,\hat{O}_2]\hat{\rho}=2\hat{O}_1\hat{\rho} \hat{O}_2-\hat{O}_1\hat{O}_2\hat{\rho}-\hat{\rho} \hat{O}_1\hat{O}_2$ and (in the second line) $\hat{O}= \tilde{\lambda}_-\hat{\sigma}_--\tilde{\lambda}_+\hat{\sigma}_+ $. Since the first line of  Eq.~(\ref{dissipator_complete}) is  dominant in the limit of large $\eta$ and $\hat{H}_{np}$ is diagonal in $\hat{\sigma}_z$, we consider the effect of the remaining terms on a density matrix with the following form:
\begin{equation}\label{simple_rho}
\hat{\widetilde{\rho}} = 
P_+ \hat\rho_{c,\uparrow} |\uparrow\rangle\langle\uparrow| +
P_- \hat\rho_{c,\downarrow} |\downarrow\rangle\langle\downarrow| ,
\end{equation}
where $\hat\rho_{c,\sigma}$ are spin-resolved density matrices for the cavity mode. 

Formally, the second line of Eq.~(\ref{dissipator_complete}) is the dominant correction to the dissipator, as it is only suppressed by $\sqrt{\eta}$. However, the effect of applying this term to Eq.~(\ref{simple_rho}) is to induce coherence between the $\uparrow$ and $\downarrow$ subspaces. Due to the large gap in $\hat{H}_{np}$, these off-diagonal terms are highly oscillating, thus they can be neglected. As for the third line of Eq.~(\ref{dissipator_complete}), it is easily seen that it is zero on a diagonal $\hat{\widetilde{\rho}}$ like Eq.~(\ref{simple_rho}). Finally, the fourth line of Eq.~(\ref{dissipator_complete}) has an influence on the detailed form of the cavity density matrices $\hat\rho_{c,\sigma}$, but it does not introduce qualitatively new physical effects. At large $\eta$, it is negligible compared to the regular photon decay, included in the first line. We conclude that the non-Lindblad terms in the second, third and fourth lines of Eq.~(\ref{dissipator_complete}) have little effect on our results, thus are omitted in the effective master equation. On the other hand, the last term of Eq.~(\ref{dissipator_complete}) is crucial to establish the populations $P_\pm$ of the spin states, thus is included in Eq.~(\ref{eq-effectiveME-normal}). 

Finally, we briefly consider the superradiant fixed point. Due to the similarity of $\hat{U}_{np}$ and $\hat{U}_{sp}$, the transformation of the dissipator of Eq.~(\ref{meq_rhod}) can be performed in a way completely analogous to the normal phase, which leads to the effective master equation for the superradiant states Eq.~(\ref{eq-effectiveME-sp}).

\section{Effective potential and noise at the second-order critical point}\label{app:Langevin_eq}

Close to the second-order transition between the normal and superradiant phases, we base our treatment on the effective Hamiltonian in the $\hat\sigma_z =-1$ subspace, given by Eq.~(\ref{Heff_xp}). The associated Heisenberg-Langevin equations read:
\begin{align}
\frac{d\hat{x} }{dt} & =  
\omega_c \left[ \alpha_y  \hat{p} +
\frac{\tilde\lambda_y^4\hat{p}^3 + \tilde\lambda_x^2 \tilde\lambda_y^2 \{\hat{p},\hat{x}^2 \}}{\eta}  \right]- \kappa \hat x +\hat{\mathcal{F}}_+, \label{Langevin_x} \\
\frac{d\hat{p}}{dt}  & = 
-\omega_c \left[ \alpha_x  \hat{x} +
\frac{\tilde\lambda_x^4 \hat{x}^3 +\tilde\lambda_x^2 \tilde\lambda_y^2\{\hat{x},\hat{p}^2\}}{\eta}  \right]-\kappa \hat{p} -i\hat{\mathcal{F}}_-,  
\label{Langevin_p}
\end{align} 
where we defined $\{\hat{A},\hat{B}\}=\frac12(\hat{A}\hat{B}+\hat{B}\hat{A})$ and $\hat{\mathcal{F}}_\pm=(\hat{\mathcal{F}}\pm \hat{\mathcal{F}}^\dag)/\sqrt{2}$, with $\hat{\mathcal{F}}(t)$ the same noise operator entering Eq.~(\ref{eq-motion}). It is also useful to rewrite Eq.~(\ref{Langevin_x}) as follows:
\begin{equation}\label{p_expression}
 \hat{p} =\frac{ \frac{d\hat{x} }{dt}+\kappa \hat x - \hat{\mathcal{F}}_+}{\omega_c\alpha_y} - 
\frac{\tilde\lambda_y^4\hat{p}^3 + \tilde\lambda_x^2 \tilde\lambda_y^2 \{\hat{p},\hat{x}^2 \}}{\eta \alpha_y} .
\end{equation}
Next, we compute $d^2\hat{x}/dt^2$ from Eq.~(\ref{Langevin_x}), leading to:
\begin{align}\label{dxd2_intermediate}
\frac{d^2\hat{x} }{dt^2}  = & -\omega_c^2 \alpha^2 \hat{x}
-2\kappa \frac{d\hat{x} }{dt}+ \mathcal{G} \nonumber \\
& -\frac{\omega_c^2}{\eta}\alpha_y \left( \tilde\lambda_x^4 \hat{x}^3 
+ \tilde\lambda_x^2\tilde\lambda_y^2 \{ \hat{x},\hat{p}^2\} \right) \nonumber \\
& +\frac{\kappa\omega_c}{\eta}\left( \tilde\lambda_y^4 \hat{p}^3 
+ \tilde\lambda_x^2\tilde\lambda_y^2 \{ \hat{p},\hat{x}^2\}\right) \nonumber \\
& + \frac{\omega_c}{\eta} \frac{d}{dt} \left( \tilde\lambda_y^4 \hat{p}^3 
+ \tilde\lambda_x^2\tilde\lambda_y^2 \{ \hat{p},\hat{x}^2\} \right) ,
\end{align}
where in the first line $\alpha^2$ is given by Eq.~(\ref{alpha_beta}) and the noise operator is:
\begin{equation}\label{G_noise}
\hat{\mathcal{G}}(t) = \frac{1}{\sqrt{2}}\left[(\kappa - i \alpha_-\omega_c)\hat{\mathcal{F}}(t) +\frac{d\hat{\mathcal{F}}(t) }{dt} + {\rm H.c.} \right].
\end{equation}
Note that, to obtain Eq.~(\ref{dxd2_intermediate}) we have substituted $d\hat{p}/dt$ with Eq.~(\ref{Langevin_p}), where $\hat{p}$ is given by Eq.~(\ref{p_expression}).

We should now simplify the nonlinear terms of Eq.~(\ref{dxd2_intermediate}). In the second and third lines, we could in principle substitute the full expression of $\hat{p}$, given by Eq.~(\ref{p_expression}). However, doing this leads to a number of uninteresting corrections. Firstly, we can safely neglect in Eq.~(\ref{dxd2_intermediate}) the higher-order terms, $\propto 1/\eta^2$, thus the $1/\eta$ term of Eq.~(\ref{p_expression}) can be omitted. Furthermore, Eq.~(\ref{p_expression}) would induce $1/\eta$ corrections to the damping and noise terms, appearing in the first line of Eq.~(\ref{dxd2_intermediate}). These effects are not important when $\eta \to \infty$. We conclude that, to deal with the nonlinear terms, we can effectively simplify Eq.~(\ref{p_expression}) to $\hat{p} \simeq  \kappa \hat{x}/(\omega_c\alpha_y )$. After this substitution, and omitting the last line of Eq.~(\ref{dxd2_intermediate}), we obtain Eq.~(\ref{damped_oscillator_quantum}) of the main text, which for convenience we repeat here:
\begin{equation}\label{damped_oscillator_appendix}
\frac{d^2\hat{x}}{dt^2}= -\omega_c^2 \left(\alpha^2 \hat x +\frac{\beta^3}{\eta} \hat{x}^3\right)-2\kappa \frac{d\hat{x}}{dt}+\hat{\mathcal{G}}.
\end{equation} 
The coefficient $\beta^2$, obtained from the second and third lines of Eq.~(\ref{dxd2_intermediate}), is given in Eq.~(\ref{alpha_beta}). Finally, we justify why the fourth line Eq.~(\ref{dxd2_intermediate}) was omitted. Due to the time derivative, this contribution will give two types of terms. The ones proportional to $d\hat{x}/dt$, following our previous arguments, can be neglected in comparison to the dominant damping term. The second type of terms is $\propto d{\hat p}/dt$, and we should express them using Eq.~(\ref{Langevin_p}). However, to obtain the effective potential we can neglect the $1/\eta$ and noise contributions, thus Eq.~(\ref{Langevin_p}) simplifies to $d{\hat p}/dt \simeq \omega_c\alpha_x \hat{x} -\kappa \hat{p}$ and, eventually:
\begin{equation}
\frac{d\hat{p}}{dt} \simeq \omega_c\alpha_x \hat{x} 
-\kappa \left(\frac{\kappa \hat{x}}{\omega_c\alpha_y }\right) 
= \frac{\omega_c}{\alpha_y}\alpha^2 \hat{x}.
\end{equation}
In conclusion, we find that, within these approximations, $d{\hat p}/dt$ vanishes at the critical point (where $\alpha^2 =0$), thus the corresponding contribution to the effective potential can be omitted.

We conclude this section by deriving the effective temperature of the bath, obtained by a comparison to the following equation of motion of a Brownian particle (we set $k_{\rm B} =1$):
\begin{equation}\label{Brownian_particle}
m \frac{d^2 y}{dt^2}= -\frac{d V(y)}{dy}-\gamma  \frac{d^2 y}{dt^2} +\sqrt{2\gamma T} \xi,
\end{equation}
where $m$ is the mass, $V(y)$ is the potential, and $\gamma$ is the damping coefficient. The noise term satisfies $\langle\xi(t)\xi(t')\rangle =\delta(t-t')$ and $\langle\xi(t)\rangle=0$. Since here $y$ is a position and $\hat {x}$ in Eq.~(\ref{damped_oscillator_appendix}) is dimensionless we define $x= y/y_0$, where $y_0$ is an arbitrary conversion length. Equivalently, we choose an energy scale $\omega_0$ giving $y_0^{-1}= \sqrt{\omega_0 m}$. First, comparing Eq.~(\ref{Brownian_particle}) to the classical limit of Eq.~(\ref{damped_oscillator_appendix}), we find the potential in the $x$ coordinate:
\begin{align}
V(x) = \frac{\omega_c^2}{\omega_0}\left(\frac12 \alpha^2 x^2 +\frac{1}{4\eta}\beta^3 x^4 \right).
\end{align}
Furthermore, proceeding as in Ref.~\cite{dalla2013keldysh}, we compute the symmetrized correlator of the noise in Fourier space:
\begin{equation}
\frac12 \langle \hat{\mathcal{G}}_\omega\hat{\mathcal{G}}_{\omega'}+\hat{\mathcal{G}}_{\omega'}\hat{\mathcal{G}}_\omega\rangle = \kappa (\kappa^2 +\alpha^2_y \omega_c^2 +\omega^2)\delta(\omega+\omega'),
\end{equation}
where $\hat{\mathcal{G}}_\omega = \int \frac{dt}{\sqrt{2\pi}} \hat{\mathcal{G}}(t) e^{i\omega t}$. The analogous correlator of the classical noise is $\frac{\sqrt{2\gamma T}}{m y_0}\delta(\omega+ \omega')$. Considering the low-frequency limit and using $\gamma/m = 2\kappa$, we finally identify:
\begin{equation}
T= \frac{\kappa^2+\alpha^2_y\omega_c^2}{4\omega_0}.
\end{equation}
We see that the choice of the energy scale is arbitrary, as $\omega_0$ simplifies when computing $V(x)/T$. The effective temperature of Ref.~\cite{dalla2013keldysh} is recovered in the isotropic limit (giving $\alpha_y=1$) and with the natural choice $\omega_0 = \omega_c$, which we assume in the main text.

\bibliographystyle{apsrev4-2}
\bibliography{AQRM}

\end{document}